\documentclass[fleqn,usenatbib]{mnras}

\usepackage{newtxtext,newtxmath}

\usepackage[T1]{fontenc}
\usepackage{calrsfs}
\DeclareMathAlphabet{\pazocal}{OMS}{zplm}{m}{n}

\DeclareRobustCommand{\VAN}[3]{#2}
\let\VANthebibliography\thebibliography
\def\thebibliography{\DeclareRobustCommand{\VAN}[3]{##3}\VANthebibliography}

\usepackage{graphicx}	
\usepackage{amsmath}	
\usepackage{comment}
\usepackage{soul}

\newcommand{\Fig}[1]{Fig.~\ref{#1}}
\newcommand{\Sec}[1]{Section~\ref{#1}}
\newcommand{\App}[1]{Appendix~\ref{#1}}
\newcommand{\Tab}[1]{Table~\ref{#1}}
\newcommand{\threehun}{{\sc The Three Hundred}}
\newcommand{\tth}{{\sc The300}}
\newcommand{\radrange}{{$[R_{200},2R_{200}]$}}
\newcommand{\calP}{\pazocal{P}}
\newcommand{\calC}{\pazocal{C}}
\newcommand{\HI}{H\,{\sc{i}}}
\newcommand{\webappurl}{https://backsplash-1050827052800.us-central1.run.app}

\defcitealias{rhee2017}{R17}

\newcommand{\perfBasicAc}{74}
\newcommand{\perfBasicBP}{59}
\newcommand{\perfBasicBC}{61}
\newcommand{\perfBasicIP}{81}
\newcommand{\perfBasicIC}{80}
\newcommand{\perfBasicTh}{0.61}

\newcommand{\perfMorphAc}{76}
\newcommand{\perfMorphBP}{61}
\newcommand{\perfMorphBC}{65}
\newcommand{\perfMorphIP}{83}
\newcommand{\perfMorphIC}{81}
\newcommand{\perfMorphTh}{0.59}

\newcommand{\perfNeighAc}{75}
\newcommand{\perfNeighBP}{61}
\newcommand{\perfNeighBC}{63}
\newcommand{\perfNeighIP}{82}
\newcommand{\perfNeighIC}{81}
\newcommand{\perfNeighTh}{0.61}

\newcommand{\perfVelocAc}{81}
\newcommand{\perfVelocBP}{69}
\newcommand{\perfVelocBC}{72}
\newcommand{\perfVelocIP}{86}
\newcommand{\perfVelocIC}{85}
\newcommand{\perfVelocTh}{0.59}

\newcommand{\perfComplAc}{84}
\newcommand{\perfComplBP}{75}
\newcommand{\perfComplBC}{75}
\newcommand{\perfComplIP}{88}
\newcommand{\perfComplIC}{88}
\newcommand{\perfComplTh}{0.58}

\newcommand{\perfEgrheAc}{72}
\newcommand{\perfEgrheBP}{61}
\newcommand{\perfEgrheBC}{40}
\newcommand{\perfEgrheIP}{76}
\newcommand{\perfEgrheIC}{88}

\newcommand{\perfEgferAc}{48}
\newcommand{\perfEgferBP}{38}
\newcommand{\perfEgferBC}{93}
\newcommand{\perfEgferIP}{90}
\newcommand{\perfEgferIC}{27}

\title[Backsplash galaxies with machine learning]{Identifying backsplash galaxies using machine learning}

\author[R. Haggar et al.]{Roan Haggar,$^{1,2}$\thanks{E-mail: rhaggar@uwaterloo.ca}
Elizaveta Sazonova,$^{1,2}$ 
Cameron R. Morgan,$^{1,2}$ 
Alexander Knebe,$^{3,4,5}$ 
Rhys Jordan,$^{6}$
\newauthor{Weiguang Cui,$^{3,4}$
Frazer R. Pearce,$^{6}$
James E. Taylor$^{1,2}$ }
\\
$^{1}$Department of Physics and Astronomy, University of Waterloo, Waterloo, Ontario N2L 3G1, Canada\\
$^{2}$Waterloo Centre for Astrophysics, University of Waterloo, Waterloo, Ontario N2L 3G1, Canada\\
$^{3}$Departamento de Fısica Teórica, M\'{o}dulo 15 Universidad Aut\'{o}noma de Madrid, E-28049 Madrid, Spain\\
$^{4}$ Centro de Investigaci\'on Avanzada en F\'isica Fundamental (CIAFF), Universidad Aut\'onoma de Madrid, 28049 Madrid, Spain\\
$^{5}$International Centre for Radio Astronomy Research, University of Western Australia, 35 Stirling Highway, Crawley, Western Australia 6009, Australia\\
$^{6}$School of Physics \& Astronomy, University of Nottingham, Nottingham, NG7 2RD, UK
}

\date{Accepted XXX. Received YYY; in original form ZZZ}

\pubyear{2026}

\begin{document}
\label{firstpage}
\pagerange{\pageref{firstpage}--\pageref{lastpage}}
\maketitle

\begin{abstract}
The galaxy population in the outskirts of a cluster contains members that have been pre-processed in groups and filaments, as well as backsplash galaxies -- those that have recently passed through the cluster's center. However, disentangling these two pathways is challenging observationally. In this work, we present a machine-learning-powered model, trained on simulations of galaxy clusters from \threehun\ suite of simulations, which can identify individual backsplash galaxies in astronomical observations. This model can build samples of backsplash galaxies with a purity and completeness of up to $\sim70\%$, and galaxies on their first infall with a purity and completeness of over $80\%$. It can be tuned to optimise either of these two metrics, and can be used with any combination of a set of observable quantities. We have also applied this model to galaxies with asymmetric \HI\ distributions in the Virgo Cluster, and have demonstrated that these galaxies are all likely approaching the cluster for the first time. This supports the idea that cold gas is removed from these galaxies soon after entering a cluster, and demonstrates how this classifier can provide a better understanding of which properties of galaxies are caused by a previous passage through a cluster. We have made this model publicly available in the form of a web app, with a link in the Conclusions of this paper.
\end{abstract}




\section{Introduction}
\label{sec:introduction}

It is well-established in the field of galaxy evolution that there exists a strong connection between the properties of galaxies, and the cosmic environments in which they are found. Early studies of the connection between galaxy properties and environment, such as \citet{oemler1974} and \citet{dressler1980}, showed that higher-density environments had a greater fraction of elliptical and lenticular galaxies, and fewer spiral galaxies. Subsequent work also showed that galaxies in dense regions of the Universe are forming fewer stars, and thus have significantly redder colours than those in sparser regions \citep{butcher1984, baldry2006, peng2010}.

Galaxy clusters are the most extreme example of a cosmic environment. These structures consist of a dark matter halo, typically of mass between $10^{14}-10^{15.5}~M_{\odot}$, containing hundreds to thousands of galaxies. The number density of galaxies in clusters is hundreds of times greater than the average density of the Universe, meaning that interactions between galaxies are frequent. Mechanisms such as tidal interactions \citep{valluri1993,smith2016,khalid2025} and harassment \citep{moore1996} can cause the morphology of cluster members to be drastically altered on relatively short timescales, although galaxy-galaxy mergers are thought to be uncommon due to the large velocity dispersion inside clusters \citep{ghigna1998, mihos2004, perez2009}. Additionally, clusters contain a hot intracluster medium, which can cause gas to be removed from galaxies through processes such as ram pressure stripping \citep{gunn1972}. Consequently, clusters have substantially different populations of galaxies than the cosmic field. Cluster galaxies typically have lower star formation rates \citep{balogh1999,lewis2002,gonzalezdelgado2022} and more bulge-dominated morphologies \citep{kuutma2017,nandi2026}. There is also evidence that the prevalence of active galactic nuclei (AGN) is different in clusters, although this relationship appears to be complex and redshift-dependent \citep{argudofernandez2016, krishnan2020, shah2025}.

Galaxy clusters grow by accretion, with galaxies falling into clusters from the field. While we might naively expect the strength of environmental processes to simply increase as we consider galaxies nearer to a cluster, the picture is actually more complex. This is due to `pre-processing': environmental processes that galaxies can experience before becoming virialised members of a cluster. Galaxies may join a cluster as individual objects, but approximately half of the galaxies in clusters are thought to have been accreted as members of a galaxy group \citep{mcgee2009, han2018}. These group environments can remove gas and quench star formation in galaxies \citep{dressler2013, pallero2019}, and enhance galaxy merger rates \citep{jian2012, benavides2020} before a galaxy reaches a cluster. Galaxies can also be accreted via cosmic filaments, which can strip gas from galaxies \citep{bahe2013, veronica2022} and so further contribute to this pre-processing. For example, \citet{nandi2026} show that the quenched fraction of galaxies becomes progressively greater from cosmic sheets, to filaments, to clusters. The overlap of different pre-processing environments -- such as galaxy groups that are also located in filaments \citep{kuchner2022, zakharova2026} -- adds further complexity to this picture. For example, \citet{kotecha2022} show that cluster member galaxies can actually be `shielded' by cosmic filaments, impeding the gas removal and star formation quenching in these galaxies, relative to other cluster members. 

A related concept to pre-processed galaxies are `backsplash galaxies'. These are galaxies that are located beyond the radius of a galaxy cluster at the present day, but have previous passed within this radius at some point in the past \citep{balogh2000, mamon2004, gill2005}. There is not an absolute consensus on how this radius should be chosen, but we use $R_{200\textrm{c}}$\footnote{$R_{200\textrm{c}}$ is defined such that the mean density within this radius is 200 times the critical density of the Universe at that redshift; we use $R_{200\textrm{c}}$ as the outer cluster radius throughout this work, and we hereafter abbreviate it to $R_{200}$.} throughout this study, for consistency with much of the existing literature. Backsplash galaxies are challenging to identify observationally, as we typically cannot determine the 3D orbits of galaxies. One notable exception to this is in the Local Group, where \citet{bennet2025} made constraints on the orbits of dwarf galaxies by measuring their proper motions, allowing a direct search for galaxies `backsplashing' from the Milky Way or M31.

As they are difficult to identify, backsplash galaxies can be considered a `contaminant' in studies of `infalling' galaxies (i.e. those on their first infall to the cluster) and pre-processing. Quenched, evolved galaxies in the outskirts of clusters would often be attributed to pre-processing. However, they may in fact be backsplash galaxies, whose properties have been impacted by a passage through the cluster centre in the last $1-3\,\textrm{Gyr}$ \citep{haggar2020}. For example, \citet{stephenson2025} find that the quenched fraction of galaxies around a sample of 11 galaxy clusters is different along the major and minor axes of the cluster, which they partially attribute to backsplash galaxies. Simulations support the idea of backsplash galaxies contaminating samples of infalling galaxies. \citet{borrow2023} use the {\sc{IllustrisTNG}} simulation \citep{nelson2019} to show that backsplash galaxies are gas-poor and have redder colours than galaxies approaching a cluster for the first time. \citet{hough2023} find similar results, using \threehun\ simulations and the semi-analytic model {\sc{sag}} \citep{cora2018}.

To disentangle the impact of clusters and pre-processing on galaxy evolution, there is interest in being able to reliably identify backsplash galaxies in observations. Most studies in this area make use of cosmological simulations. In \citet{haggar2020}, we showed that backsplash galaxies are more prevalent around centrally-dominated, dynamically relaxed galaxy clusters. Fewer backsplash galaxies exist around disturbed clusters, and they tend to be concentrated closer to $R_{200}$. Other studies have found a similar connection between the related `splashback radius', and the formation histories of galaxy clusters \citep[e.g.][]{more2015, adhikari2021, haggar2024_splashback}. These findings allow the contamination from backsplash galaxies to be quantified on a cluster-by-cluster basis. Multiple other studies have gone beyond this statistical approach, by separating backsplash galaxies from infalling galaxies based on their positions in the projected cluster-centric position-velocity phase space. \citet{rhee2017} and \citet{martinez2023} both identify galaxies close to $R_{200}$ ($d_{\textrm{proj}}\lesssim1.5R_{200}$) and with low line-of-sight velocities ($v_{\textrm{LOS}}\lesssim0.5\sigma$) as backsplash galaxies, consistent with them being at the apocentres of their orbits. Other studies have improved on this approach; \citet{delosrios2021} developed a machine-learning-powered approach to reconstruct galaxies' orbits based on this phase-space information, to separate backsplash galaxies. Other properties of galaxies are also promising avenues for identifying backsplash galaxies. For example, \citet{wang2024} suggest that backsplash early-type galaxies could be identified as outliers from the X-ray luminosity-temperature relation, due to the heating of their gas during a recent passage through a cluster.

In this study we aim to improve on these existing approaches, by using machine learning methods to identify galaxies that have previously passed through a cluster on a case-by-case basis, but using properties beyond their phase-space information. Specifically, we use a random forest classifier, trained on \threehun\ suite of hydrodynamical simulations of large galaxy clusters \citep[hereafter \tth;][]{cui2018,cui2022}. We train this model on observable properties of galaxies extracted from the simulations, combined with a measure of the dynamical state of galaxy clusters. This approach allows us to identify individual backsplash galaxies, and to separate galaxies into two populations (infallers and backsplash), that are no longer separated by a hard boundary in position-velocity phase space. The model is designed in such a way that a user can choose any number from a variety of input parameters to classify a sample of galaxies. Additionally, by tuning the classification threshold, this model can be used to build either highly-complete or highly-pure samples of either of these two populations, which will be crucial in making robust conclusions about the impact of a cluster on galaxy properties. We have also made this classifier available to the astronomy community, as a live web app (with detailed user instructions in \App{sec:web_app}).

The structure of the paper is as follows: in \Sec{sec:methodology}, we describe our methodology, including the simulations (\Sec{sec:simulations}) and machine learning methods (\Sec{sec:rf}) we use, and the possible input parameters for the classifier (\Sec{sec:properties}). We then present our results in \Sec{sec:results}, evaluating the performance of the model with a variety of input parameters. In \Sec{sec:applications} we describe the applications of this model, including applying it to a sample of ram pressure stripping candidate galaxies in the Virgo Cluster (\Sec{sec:virgo}). Finally, we summarise the paper in \Sec{sec:conclusions}.

\section{Data and Methodology}
\label{sec:methodology}

\subsection{Simulations}
\label{sec:simulations}

This work uses data from \threehun\ project \citep{cui2018}, a suite of hydrodynamical zoom resimulations of large galaxy clusters. The simulation suite is based on the MultiDark Planck 2 simulation \citep[MDPL2; ][]{klypin2016}\footnote{The MultiDark simulations are publicly available from the CosmoSim database, \url{https://www.cosmosim.org}.}, a dark matter-only simulation with a comoving box size of $1~h^{-1}\rm{Gpc}$. \tth simulations use the same $\Lambda\rm{CDM}$ cosmology as MDPL2 \citep[$\Omega_{\rm{M}}=0.307$, $\Omega_{\rm{B}}=0.048$, $\Omega_{\Lambda}=0.693$, $h=0.678$, $\sigma_{8}=0.823$, $n_{\rm{s}}=0.96$;][]{planck2016}\footnote{The reduced Hubble constant, $h$, is defined such that the Hubble constant, \mbox{$H_{\mathrm{0}}=h\times100~\mathrm{km}\,\mathrm{s}^{-1}\,\mathrm{Mpc}^{-1}$}.}.

To produce \tth simulations, the 324 most massive dark matter haloes from the MDPL2 simulation were selected at $z=0$. For each of these, all particles from within $15~h^{-1}\,\mathrm{Mpc}$ of the halo's centre were traced back to their initial conditions, at $z_{\mathrm{init}}=120$. These dark matter particles were split into a dark matter particle and a gas particle, with respective masses determined by the cosmic baryon fraction ($m_{\mathrm{DM}}=1.27\times10^{9}~h^{-1}\,M_{\odot}$ and $m_{\mathrm{gas}}=2.36\times10^{8}~h^{-1}\,M_{\odot}$). Dark matter particles outside of this central region at $z=0$ were degraded in resolution to reduce computing time. Each of these clusters was then resimulated using full baryonic physics, to produce 324 hydrodynamical simulations of galaxy clusters, each embedded in a lower-resolution dark matter-only cosmological box.

These simulations have been run with multiple different physics models using the same initial conditions, but in this work, we utilise the {\sc{Gizmo-Simba}} run \citep{cui2022}, which uses the {\sc{Gizmo}} hydrodynamical code \citep{hopkins2015}, and the galaxy formation models of the {\sc{Simba}} simulation \citep{dave2019}. The {\sc{Gizmo}} code uses a meshless finite mass solver to evolve the gas component of the simulated clusters, while accounting for shear, shocks, and viscosity. The {\sc{Simba}} galaxy formation code uses {\sc{Gizmo}}, and is an improved version of the {\sc{Mufasa}} model \citep{dave2016}. Radiative cooling and heating of gas are implemented using the {\sc{Grackle-3.1}} library \citep{smith2017}, which gives improved thermal baryonic properties due to its `on-the-fly' self-shielding \citep[see][]{rahmati2013}.  The $\mathrm{H}_{2}$-based star formation is based on the \citet{krumholz2011} prescription, where star formation is triggered at a specific $\mathrm{H}_{2}$ density and metallicity. Stellar feedback from supernovae and star formation is carried by two-phase winds. Black holes are seeded in galaxies with a stellar mass greater than $10^{10.5}~h^{-1}\,M_{\odot}$, a factor of 10 greater than the original {\sc{Simba}} model, due to the lower resolution in \tth. AGN feedback consists of both radiative and jet modes. A far more extensive description of the implementation of {\sc{Gizmo}} and {\sc{Simba}} in \tth can be found in \citet{cui2022}. The full simulation suite consists of 324 large galaxy clusters, ranging in mass\footnote{Throughout this work, the cluster mass definition we use is $M_{200}$, which is the mass enclosed by a sphere of radius $R_{200}$.} from $5\times10^{14}~h^{-1}\,M_{\odot}$ to $2.6\times10^{15}~h^{-1}\,M_{\odot}$, corresponding to cluster radii, $R_{200}$, of between $1.3-2.2~h^{-1}\,{\mathrm{Mpc}}$.

This work is designed to be applicable to massive, low-redshift clusters. For that reason, we restrict our analysis to the final snapshot of the simulations, at $z=0$. However, we use the full $z>0$ merger trees to construct catalogues of infalling and backsplash galaxies. We plan to extend this analysis to higher-redshift clusters in the future.

\subsubsection{Galaxy identification and tree-building}
\label{sec:galaxy_identification}

The dark matter haloes and subhaloes in \tth are identified using the Amiga Halo Finder, {\sc{ahf}}\footnote{\url{http://popia.ft.uam.es/AHF}} \citep[see][]{gill2004_ahf, knollmann2009}. {\sc{ahf}} is a density peak halo finder, and is used to associate particles to the main, central cluster halo, as well as identify cluster subhaloes, and haloes outside of the cluster. {\sc{ahf}} was run at each of the 129 saved redshifts of \tth clusters, and these catalogues were then linked together using {\sc{ahf}}'s in-house tree-builder, {\sc{mergertree}}. For each halo at a given redshift, this tree-builder searches through all prior snapshots for a main progenitor\footnote{The main progenitor of a halo is chosen to maximise the merit function implemented by {\sc{mergertree}}, given by Eq. B1 in \citet{knebe2013}.}, as well as secondary progenitors that merge into the main branch of the merger tree \citep[see][for additional details on {\sc{mergertree}}]{knebe2011_halo_comparison, srisawat2013}. Throughout this work, we only consider galaxy haloes with total masses $m_{200}>10^{10.5}~h^{-1}\,M_{\odot}$, corresponding to approximately 100 particles. We also place a stellar mass cut, excluding all galaxies with $m_{\mathrm{star}}<10^{9.5}~h^{-1}\,M_{\odot}$. Galaxy R-band magnitudes are used in calculating $M_{12}$, the magnitude difference of the brightest two cluster members. These magnitudes are calculated with the {\sc{Python}} package {\sc{Caesar}}\footnote{\url{https://caesar.readthedocs.io/en/latest}}, which uses the light from stars within $30~\mathrm{kpc}$ of each galaxy's centre.

\subsection{Backsplash galaxies}
\label{sec:backsplash}

The focus of this work is on backsplash galaxies. These galaxies reside outside of a cluster's radius ($R_{200}$) at the present day, in the `infall region', typically associated with galaxies falling into the cluster for the first time. However, backsplash galaxies have passed within $R_{200}$ at some time in the past, travelling through the cluster centre which contains the virialised population of cluster member galaxies. Throughout this work, in both simulations and observations, we define backsplash galaxies in 3D, not in 2D projected coordinates. This means that some backsplash galaxies can appear to lie inside $R_{200}$ at the present day, as they are along the same line-of-sight as the cluster halo. 

Although we are defining backsplash galaxies in 3D, we focus this study only on the projected annulus between \radrange; we exclude the projected region within $R_{200}$ of a cluster, which contains backsplash galaxies, infalling galaxies, and cluster member galaxies. This full classification of bound, infalling, and backsplash galaxies is a more complex problem, although it is one we plan to tackle in a future study. By considering only the annulus between \radrange, some backsplash galaxies will be missed if they exist along the same line-of-sight as the cluster. We elaborate on this at the end of \Sec{sec:rf}. 

As mentioned in \Sec{sec:introduction}, some previous studies have defined backsplash galaxies differently to this. Some early studies \citep{mamon2004, gill2005} select galaxies beyond $R_{\rm{vir}}\sim R_{100}\sim 1.4R_{200}$, that have passed within this distance during their pericentric passage. Our definition, using $R_{200}$ as a boundary, is now fairly widely used, but is still a somewhat arbitrary choice. Selecting galaxies that have `backsplashed' from within the splashback radius \citep{more2015, adhikari2021}, or that have previously passed through a cluster's accretion shock radius, are perhaps more physically-motivated choices; \citet{vijayaraghavan2013} and \citet{zhang2021} both show that ram pressure is enhanced when galaxies pass through an accretion shock. However, we use the standard definition, based on $R_{200}$, for two main reasons. Firstly, most galaxies enter a cluster on radial orbits, meaning their pericentric distances are small. In our simulations, $91\%$ of backsplash galaxies have passed within $0.7R_{200}$ of the cluster centre. Similarly, \citet{gill2005} find that $90\%$ of backsplash galaxies have previously passed within $0.5R_{\rm{vir}}\simeq0.7R_{200}$, while \citet{wetzel2014} show that $80\%$ of backsplash galaxies outside of $R_{\rm{200m}}\sim1.6R_{200}$ were previously inside $0.8R_{200}$. These small pericentres mean that there are few backsplash galaxies that are `missed' by not choosing a larger boundary, such as the splashback radius or accretion shock. Secondly, the number of backsplash galaxies drops sharply with cluster-centric distance; fewer than $10\%$ of backsplash galaxies reside beyond $2R_{200}$ from a cluster \citep{haggar2020, borrow2023}. We therefore expect the impact of backsplash galaxies on the properties of the overall galaxy population to be small at distances far from the cluster centre, meaning it is more important to study these galaxies at lower radii, such as our chosen range, \radrange.

Our complete sample of 324 simulated galaxy clusters contain a total of 172079 galaxies in the 2D projected region \radrange\ (56505 backsplash galaxies, and 115574 infalling galaxies). Our final dataset contains three times as many galaxies, as we project each of our galaxy clusters along three orthogonal lines of sight. We use this dataset to construct our classifier.

\subsection{Random Forest Classifier}
\label{sec:rf}

Machine learning tools have become widely used in classification tasks in recent years \citep[e.g.][]{banerji2010, dominguezsanchez2018, marini2022, hao2026}. Random forest classifiers are a very commonly used example of such a tool for multiple reasons. Predominantly, this is because of their ease of interpretability, and their low computational cost, which means new models can be produced locally and on-the-fly.

These classifiers take a set of objects (`elements'), each of which has a number of `features', and assigns each object to a `class'. In our case, each galaxy in the projected annulus \radrange\ around a cluster is an `element', and the properties of this galaxy are the `features'. These include quantities such as the stellar mass of the galaxy and its cluster-centric distance; we describe these in detail in \Sec{sec:properties}. Each of the galaxies in this annulus falls into one of two classes, `backsplash' or `infalling' (i.e. `not backsplash'). 

Using the training data, the random forest classifier constructs a decision tree to separate the data into these two classes. First, the data is split into two subsets (`nodes'), by choosing a boundary value in one of the features. In our case, this value is chosen such that it minimises the weighted average Gini impurity in the two subsets, where the Gini impurity of one subset is given by:

\begin{equation}
     G=1-\sum_{i=1}^{K}p_i^2\,.
     \label{eq:gini}
\end{equation}
In our case, $K=2$ (the number of classes: `backsplash' and `infalling'), and $p_{i}$ is the proportion of each class that makes up the subset of data. Consequently, Gini is minimised when a subset only contains members from one class (e.g. $p_{1}=1$, $p_{2}=0$). Each of these two subsets is then split again based on another feature, and this process repeats until a decision tree has been constructed that separates the data into numerous subsets, each optimised for maximum purity. 

A random forest classifier then repeats this process, by sampling the full set of input elements (for example, by bootstrap sampling the training set of galaxies), and generating a new, independent tree. Finally, these trees are combined to act as an ensemble -- this large number of trees makes up the `forest'. Each object is assigned a likelihood of being a member of each class by the classifier, based on the number of trees in the ensemble that `voted' for it to belong to this class. In our implementation, each tree only trains from a subset of the overall training set, and thus learns different features. Such an ensemble of trees means that a random forest classifier is less sensitive to over-fitting and outliers in the training data, which can impact individual decision trees. The trained random forest can then be used to predict which class an element will fall into, based on its features. For an extensive description of random forest classifiers, see \citet{breiman2001}; for a concise, descriptive summary, see \citet{piotrowska2022}.

To train our decision tree, we first split the set of all galaxies into a training and a test set. We reserve all galaxies from 65 clusters as a `test' set ($20\%$ of our galaxies), which the random forest never sees. The remaining $80\%$ of galaxies are used in training. Each forest, by default, consists of 50 trees with depth 30. We chose this depth as, beyond this, the accuracy plateaus as a function of tree depth in our tests. More trees would lead to better uncertainty estimates and finer control over the classification threshold (see below), where 50 trees allows us to control the threshold within 2\%. Each tree then trains on a balanced subset of the training data, consisting of 15,000 backsplash and 15,000 infalling galaxies (this number can be changed). The number of trees, tree depth, and the subset size can be changed by the user.

Random forest classifiers are most effective at separating classes that are equally represented in a population. This is one of the reasons why we only consider the projected annulus \radrange around clusters: due to the high number density in the centres of clusters, the large majority of galaxies along the line-of-sight of the cluster centre are cluster members, with backsplash and infalling galaxies only making up a small minority. A random forest classifier is hence much better suited to classifying objects in the projected outskirts of a cluster, where the two populations of interest are much more evenly balanced. 

Additionally, we expect the properties of backsplash and infalling galaxies to be distinct, partially because these galaxies have joined the cluster at distinct times. Infalling galaxies are joining a cluster at the present day, while most backsplash galaxies entered the cluster $2-4\,\textrm{Gyr}$ ago \citep{haggar2020}. However, galaxies in the central regions of a cluster have joined continuously throughout the history of the cluster, making them more diverse in their properties. Consequently, we focus only on separating the two evenly balanced populations with distinct histories in the projected annulus \radrange.

\subsubsection{Classification threshold}
\label{sec:threshold}

A random forest classifier produces a likelihood of each object being a member of each class. Objects are then assigned to classes based on whether they fall above a `threshold' value of this measure. However, selecting different values of this threshold will produce samples with varying levels of purity, $\calP$, and completeness, $\calC$. These are defined as:
\begin{equation}
     \calP=\frac{TP}{TP+FP}\,,
     \label{eq:purity}
\end{equation}
\begin{equation}
     \calC=\frac{TP}{TP+FN}\,,
     \label{eq:completeness}
\end{equation}
where $TP$, $FP$, and $FN$ correspond to the number of true positive, false positive, and false negative classifications. A high threshold value will typically result in a sample with high purity (low number of false positives), but a low completeness (high number of false negatives). Similarly, a low threshold will give a low purity, but a high completeness. The threshold that correctly classifies the greatest number of objects is typically between these two.

As stated above, our decision trees are trained on balanced subsamples of the training set (15,000 backsplash/infalling galaxies), and in this case, a classification threshold of $p=0.5$ is the one that achieves the best purity and completeness for both classes. However, even in the annulus \radrange, there are still more than twice as many infalling galaxies as backsplash, and so a relatively high `accuracy' of $\sim67\%$ could be achieved by simply assigning all galaxies as infalling. This is a well-known problem in machine learning \citep[e.g.][]{bickley2021}.

One way to remedy this is to adjust the classification threshold, p(Backsplash), to tune to a desired purity or completeness on the overall data, not just a balanced dataset. Increasing $p$ from $50\%$ would therefore increase purity and accuracy, sacrificing completeness. In this work, we choose a threshold value such that the predicted fraction of backsplash galaxies matches the actual backsplash fraction, on a cluster-by-cluster basis, for our training dataset. In our case, this threshold value also happens to be very close to the value that would maximise the accuracy of the classifier, around $p\sim0.6$ for most iterations of the model. The threshold can be changed manually, to produce datasets with a greater purity, or a greater completeness. This represents one of the improvements that our model has over pre-existing methods of identifying backsplash galaxies. We discuss this further in \Sec{sec:results}.

\subsubsection{Properties of simulated galaxies}
\label{sec:properties}

The features used by the classifier we trained are all properties of the simulated galaxies. Crucially, every property the model was trained on is one that could be feasibly measured in observations. These range from properties that are trivial to measure, to ones that require more specialised observations; the model is trained on a subset of these, specified by the user. In all, 14 properties can be included, which are described below:

\begin{itemize}
\item $R_{200}$: Radius of the galaxy cluster, in $\rm{kpc}$, defined such that the density of the enclosed region is equal to 200 times the critical density of the Universe.
\item $d_{\rm{proj}}$: Projected 2D distance from cluster centre (given by location of the cluster's central galaxy), in units of $R_{200}$ of the cluster. All of the galaxies used in this model lie in an annulus between $R_{200}$ and $2R_{200}$.
\item $v_{\rm{LOS}}$: Line-of-sight velocity of the galaxy, relative to the cluster. Measured as a fraction of the cluster's line-of-sight velocity dispersion.
\item $M_{12}$ Magnitude difference in $R$-band between the brightest and second brightest galaxies within $R_{200}$ of a cluster's centre (projected distance). Magnitudes are calculated using light from within $30~\mathrm{kpc}$ of each galaxy's centre. We note that this is a property solely of the cluster, and so is the same for all galaxies around a given cluster. This measure was motivated by the fact that magnitude difference is a strong tracer of dynamical state \citep{ragagnin2019,haggar2024_ds,ahad2025}, and that dynamically relaxed clusters have a greater fraction of backsplash galaxies in their outskirts \citep{haggar2020}.
\item $m_{*}$: Galaxy stellar mass. This is taken directly from the simulations, rather than from mock observations, to avoid any assumptions about spectral energy distributions being built into the model. Stellar mass includes all stars identified as a part of the galaxy by the halo finder. Given in units of $M_{\odot}$.
\item $m_{\rm{h}}$: Galaxy halo mass: the $m_{200}$ mass of each galaxy. Also given in units of $M_{\odot}$. While this is very challenging to measure on a galaxy-by-galaxy basis, targeted studies have shown that it is possible to estimate this quantity \citep[e.g][]{cappellari2013, vandesande2021, ciocan2026}.
\item cos$(\theta_{\rm{v}})$: For each galaxy, the angle between the galaxy's velocity in the 2D plane of the sky, and the (2D projected) vector pointing to the cluster centre. A galaxy moving directly towards the cluster centre would have $\theta_{\rm{v}}=0^\circ$, while a galaxy moving directly away from the cluster centre would have $\theta_{\rm{v}}=180^\circ$. While this cannot be measured directly, there exist methods that allow this to be estimated -- for example, by measuring the angles of AGN jets or ram pressure stripped tails \citep{roberts2020}.
\item $d_{\rm{proj,nn4}}$: Proper projected distance of target galaxy to fourth nearest neighbour, above a stellar mass limit of $10^{9.5}~h^{-1}\,M_{\odot}$. This is given in units of $R_{200}$ of the main cluster halo, to prevent mass-dependent effects as much as possible; \citet{chamberlain2024} showed that close pairs of galaxies are best identified when the threshold distance depends on the radius of the massive halo in the pair. Backsplash galaxies are typically not found in galaxy groups \citep{haggar2023}, so we would expect infalling galaxies to have a smaller mean distance to their neighbouring galaxies than backsplash galaxies.
\item cos$(\psi_{\rm{BCG}})$: Cosine of the angle between the 2D vector pointing to a galaxy's position, and the major axis of the stellar component of the central galaxy. Cosmic filaments are more likely to exist in the direction of the major axis of a cluster's central galaxy \citep{kuchner2020, smith2023}, and so this parameter is designed to probe populations of galaxies preferentially associated with filaments.
\item cos$(\phi_{*})$: Cosine of the angle between the 2D vector pointing to cluster centre, and the major axis of the stellar component of a galaxy. The major axis was identified by calculating the reduced inertia tensor of the stellar particles of each galaxy. This parameter was motivated by works such as \citet{knebe2020}, who used simulations to show that the stellar component of backsplash galaxies is preferentially aligned towards the cluster centre. We note, however, that observational studies find mixed results on the degree of the radial alignment of galaxies to cluster centres \citep[e.g][]{huang2018,chan2021}. $\phi_{*}$ is constrained to have values between $0^\circ$ and $90^\circ$; there is no physical distinction between a galaxy's major axis being perfectly aligned ($\phi_{*}=0^\circ$) or perfectly anti-aligned ($\phi_{*}=180^\circ$) to the cluster centre, as the `positive' direction is chosen arbitrarily.
\item $e^{*}_{\rm{a}}$: The ratio between the minor and major axes of the stellar component of each galaxy. \citet{knebe2020} showed that backsplash galaxies have elongated stellar components, most likely due to tidal forces experienced during pericentric passage through the cluster.
\item $e^{*}_{\rm{a,BCG}}$: The ratio of the minor and major axes of the stellar component of the central cluster galaxy. Clusters with elongated central galaxies have elongated haloes, and so a more asymmetric distribution of galaxies \citep{niedersteostholt2010}.
\item $a_{\rm{hl}}$: Offset of centre of light in inner and outer regions of the galaxy. To find this, we calculate the projected centre-of-mass of all the stellar particles inside the half-light radius of each galaxy, $r_{\rm{hl}}$. We also calculate this for all particles between $r_{\rm{hl}}$ and $2r_{\rm{hl}}$ of the centre of each galaxy. $a_{\rm{hl}}$ is then equal to the difference between these, as a fraction of the half-light radius. This parameter is loosely equivalent to the centroid shift, $W$ \citep{mohr1993} or alignment, $A$ \citep{mantz2015}, used in calculating X-ray morphologies of galaxy clusters.
\item $b_{\rm{hl}}$: Ratio between axis ratio of stars inside $r_{\rm{hl}}$ of a galaxy, and the axis ratio of stars between $r_{\rm{hl}}$ and $2r_{\rm{hl}}$. This `ratio of ratios' distinguishes galaxies with tidal features in their outskirts, even if their central regions are undisturbed.
\end{itemize}

We note that some of these `galaxy properties' are in fact properties of the host cluster -- for example, $R_{200}$ and $M_{12}$. Because of this, all galaxies from a given cluster will have the same value for these properties. Our model is trained on data from many galaxy clusters, and so there is still a wide distribution of values for these parameters. 

Additionally, we do not use an exhaustive list of galaxy properties to train our model, and exclude multiple galaxy properties such as colours, metallicities, and gas masses. These quantities are heavily dependent on the implementation of baryonic and sub-grid physics in the simulations, and so we opt to exclude them from our classifier. Instead, we focus on simpler geometric observables or tracers of galaxies' overall stellar mass and shape. In a later version of this model, we aim to include data from different hydrodynamical simulations. This will include other iterations of \tth{} simulations, that will simulate the same galaxy clusters as in this work, but using hydrodynamical/galaxy formation codes beyond the {\sc{Gizmo-Simba}} codes. We will then be able to test the robustness of our model against baryonic physics, and potentially include more measures of galaxy properties.

The goal of our algorithm is for a user to obtain galaxy classifications, using only the subset of these parameters they have available. To do this exhaustively would involve training $2^{14}-1\approx10^{4}$ different models, for each possible combination of parameters. Instead, we built a web app that retrains a new model on-the-fly each time a user selects a new subset of parameters. Since random forests are computationally lightweight, this takes only a few minutes of computational time. In this work, we train the random forest classifier with several different combinations of these 14 parameters, to test the efficacy of this model when applied to data typically available from different types of astronomical observations. Initially, in \Sec{sec:complete}, we test the model using all 14 of the parameters described above. We then re-test the model using several subsets of the parameters, which represent the data that would be more realistically available for a large sample of galaxies. Specifically, the subsets of parameters that we investigated are:

\begin{itemize}
\item $d_{\rm{proj}}$, $v_{\rm{LOS}}$, $M_{12}$: A selection of basic dynamical properties that would be available from spectroscopic measurements of cluster members.

\item $d_{\rm{proj}}$, $v_{\rm{LOS}}$, $M_{12}$, $\theta_{\rm{v}}$: Basic three properties, plus the direction of velocity in the plane of the sky, which can be inferred from the angles of AGN jets or ram pressure stripped tails.

\item $d_{\rm{proj}}$, $v_{\rm{LOS}}$, $M_{12}$, $d_{\rm{proj,nn4}}$, cos$(\phi_{*})$, $e^{*}_{\rm{a,BCG}}$: Basic three properties, plus distance to fourth nearest neighour, central galaxy shape, and position relative to the central's major axis. This information would be available from wide-field mass-complete spectroscopic observations of cluster members, such as the WEAVE Wide-Field Cluster Survey \citep[WWFCS;][]{jin2024}, and the CHileAN Cluster galaxy Evolution Survey \citep[CHANCES;][]{sifon2025}.

\item $d_{\rm{proj}}$, $v_{\rm{LOS}}$, $M_{12}$, $m_{*}$, cos$(\phi_{*})$, $e^{*}_{\rm{a}}$, $a_{\rm{hl}}$, $b_{\rm{hl}}$: Basic three properties, plus stellar mass and morphological parameters of the galaxies.
\end{itemize}

Although these are the combinations of parameters that the model was tested with, we note that any combination of these can be provided by a user, depending on the information they have available. For example, the combination of parameters we use in \Sec{sec:virgo} does not exactly match any of the combinations listed above.

We also note that there are some quantities that are implicitly used as inputs. For example, the distances of galaxies from the cluster centre are given in units of the cluster's $R_{200}$ radius, and the velocities of galaxies in units of the cluster velocity dispersion. Consequently, we assume that these basic properties of a cluster are also known when using this model.

\section{Results}
\label{sec:results}

\begin{figure}
\includegraphics[width=\columnwidth]{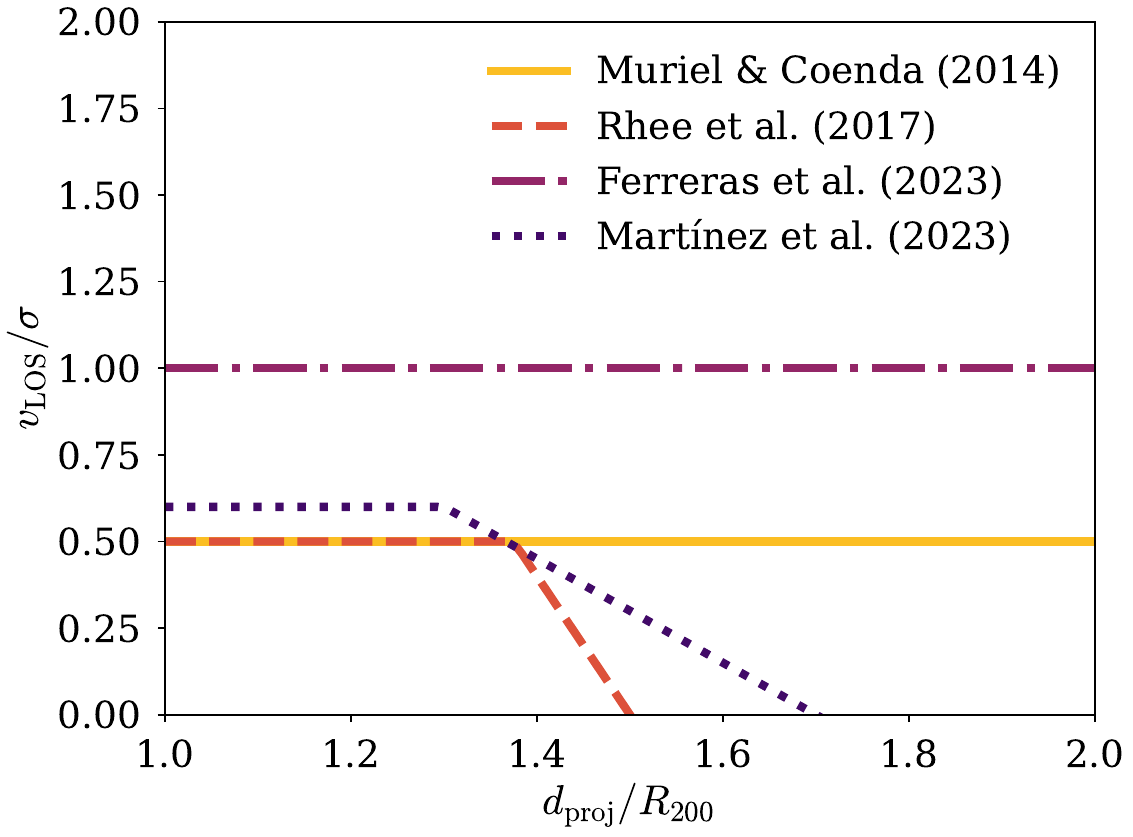}
\caption{Plot showing region of cluster-centric phase space stated to contain backsplash galaxies, from a selection of previous works in the literature \citep{muriel2014, rhee2017, ferreras2023, martinez2023}. In each case, the `backsplash region' is the region below these lines. In some of these cases \citep{rhee2017, ferreras2023}, the backsplash region extends to radii outside of the range \radrange, but as our work only studies this range, we only plot the constraints within this radial range.}
\label{fig:phase_space_boundaries}
\end{figure}

In this section, we present some results from our model, evaluating its effectiveness at classifying galaxies in the outskirts of simulated clusters into backsplash galaxies and first-time infallers. We also compare our model to some previous methods. Most existing methods of classifying backsplash galaxies use projected phase-space information, and select backsplash galaxies as those residing in a particular region of the phase space defined by 2D positions and line-of-sight velocities. Multiple definitions exist in the literature of this `backsplash region'; we have illustrated a selection of these \citep{muriel2014, rhee2017, ferreras2023, martinez2023} in \Fig{fig:phase_space_boundaries}. 

There is not a clear consensus on how to select backsplash galaxies from this phase-space information, although all of these definitions have some common features. Each definition focuses on galaxies with low velocities, in the infall region of clusters (approximately around \radrange). Each definition also produces a clear boundary between the infalling and backsplash galaxies, which is essentially a necessity when only using phase-space information. Among these multiple definitions, there is not a clear answer to which definition is the `best' -- stricter definitions will exclude more backsplash galaxies, but are also likely to produce a cleaner, purer sample of these objects. We note that this is the case here: between \radrange, the backsplash galaxies selected by \citet{rhee2017} are a strict subset of those selected by both \citet{muriel2014} and \citet{martinez2023}, which are in turn both subsets of those selected by \citet{ferreras2023}. We opted to compare our results to the definition of \citet{rhee2017} (hereafter \citetalias{rhee2017}), and so throughout the rest of this work we use the \citetalias{rhee2017} classification as a baseline to evaluate our model.

While it would be preferable to verify this model against real data, a ground truth of whether or not an observed object is a backsplash galaxy is not easy to obtain. All of the verification of our model is therefore carried out by applying it to the test set of our simulated data, after training it independently on the training set. In \Sec{sec:applications}, we discuss the potential application of this model to observational data.

\subsection{Complete set of galaxy properties}
\label{sec:complete}

The first iteration of this model we tested is the complete random forest classifier. In this case, all 14 of the parameters listed in \Sec{sec:properties} are provided, to make the most accurate classification of galaxies.

\begin{figure}
\includegraphics[width=\columnwidth]{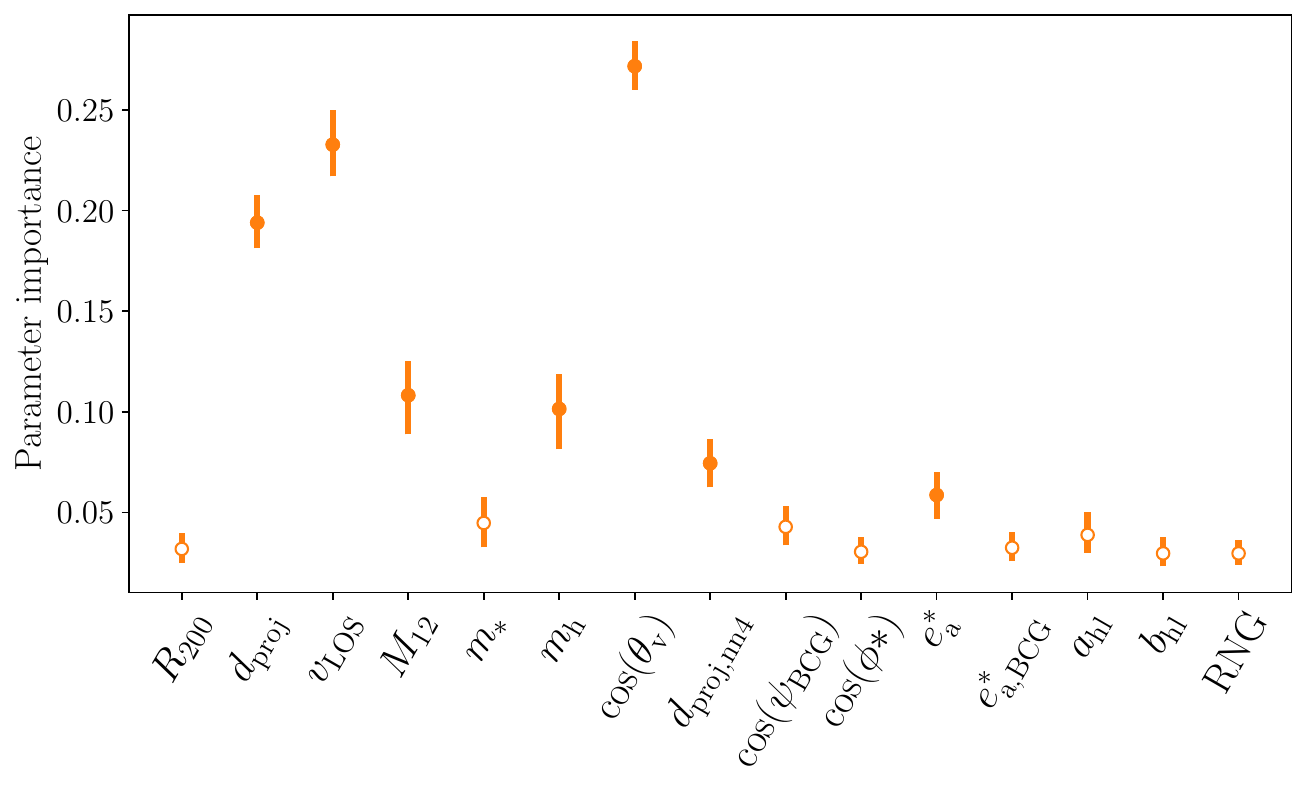}
\caption{Relative importance of all 14 measurable quantities in classifying backsplash galaxies, plus a random number generator as a baseline measurement. This is used to identify metrics that provide no classifying power, which are shown as unfilled circles.}
\label{fig:param_importance}
\end{figure}

The permutation feature importances of each of these 14 features are shown in \Fig{fig:param_importance}. These permutation importances are calculated by first running the classifier on the complete test dataset, and evaluating its performance. The values of one of the features are then randomly shuffled between all galaxies, the classifier is re-run, and the decrease in performance is measured. Repeating this process gives a thorough measure of the classifying power provided by each of the features. A user can therefore refer to \Fig{fig:param_importance} to decide which parameters would be valuable to observe and include to improve classification. Of these 14 parameters, seven of them ($R_{200}$, $m_{*}$, $\rm{cos}(\psi_{\rm{BCG}})$, $\rm{cos}(\phi{*})$, $e^{*}_{\rm{a,BCG}}$, $a_{\rm{hl}}$, and $b_{\rm{hl}}$) do not provide any substantial improvement in the classification of backsplash and infalling galaxies; these are shown as empty circles in \Fig{fig:param_importance}. In addition to these, $e^{*}_{\rm{a}}$ provides only minor classifying power, at a significance of approximately $2.1\,\sigma$. The remaining six parameters all contribute to the identification of backsplash galaxies ($>3\,\sigma$ significance).

There are various reasons for this. Due to the limited resolution of the simulations, some of these galaxy-scale properties such as $a_{\rm{hl}}$ and $e^{*}_{\rm{a}}$ are being calculated using only tens to hundreds of star particles, meaning that they are likely to be noisy. This is not the case for the cluster radius, $R_{200}$. However, we note that this quantity is already implicitly used by the classifier, as the distances $d_{\rm{proj}}$ are all normalised by the cluster radius. It is therefore not unexpected that including this parameter a second time would provide no further information to the model.

The six features that all provide a substantial amount of information when classifying galaxies are $d_{\rm{proj}}$, $v_{\rm{LOS}}$, $M_{12}$, $m_{\rm{h}}$, $\rm{cos}(\theta_{\rm{v}})$, and $d_{\rm{proj,nn4}}$. Again, it is not unexpected that any of these provide information allowing us to identify backsplash galaxies. Numerous previous studies \citep[e.g][]{muriel2014, rhee2017, ferreras2023, martinez2023} have found backsplash galaxies using the projected $v_{\rm{LOS}}-d_{\rm{proj}}$ velocity-position phase space. Including the velocity angle $\theta_{\rm{v}}$ then provides stronger constraints on the 3D velocity vector. The $R$-band magnitude difference, $M_{12}$, probes a cluster's dynamical state, and it has been previously established that relaxed clusters have more backsplash galaxies in their outskirts \citep{haggar2020}. There is also evidence that galaxies in bound groups nearby to clusters are typically first-time infallers \citep{haggar2023}, and so $d_{\rm{proj,nn4}}$ can help to quantify this. Finally, backsplash galaxies are likely to be heavily tidally stripped \citep{muldrew2011, smith2016}, so their halo masses, $m_{\rm{h}}$, are typically lower. It is important to note that measuring halo masses for individual galaxies is observationally challenging, typically requiring high-quality IFU data to allow an independent estimate of the halo mass to be made \citep[e.g][]{cappellari2013, vandesande2021, ciocan2026}. 

\begin{figure}
\includegraphics[width=\columnwidth]{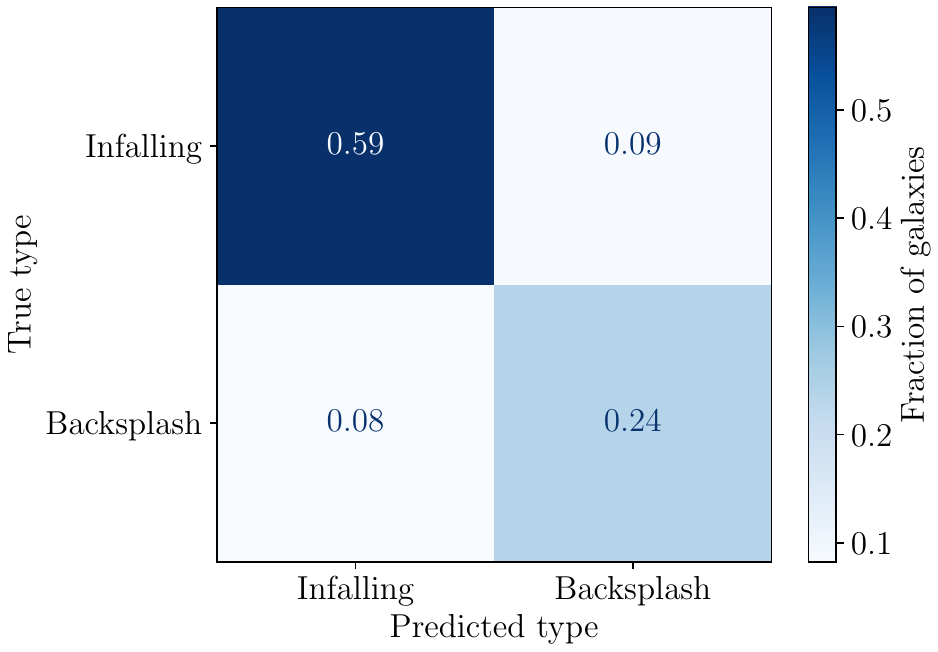}
\caption{Confusion matrix of classifier when using all 14 measurable quantities. Each panel shows fraction of galaxy sample in the test set of data that are classified correctly/incorrectly as backsplash/infalling galaxies.}
\label{fig:confusion_all}
\end{figure}

The classification of backsplash/infalling galaxies from this full model has a measured accuracy of $\perfComplAc\%$, for galaxies in the projected annulus between \radrange. The sample of backsplash galaxies that are produced has a purity of $\calP=\perfComplBP\%$ and a completeness of $\calC=\perfComplBC\%$, while the galaxies that are classed as first-time infallers have a purity and completeness of $\calP=\perfComplIP\%$ and $\calC=\perfComplIC\%$, respectively. This is illustrated in a confusion matrix, \Fig{fig:confusion_all}. If only the six `important' parameters described above are included, these values decrease by only $1-2\%$, again illustrating that little extra information is provided by these additional quantities. We do not find a significant dependence of the model accuracy on cluster mass (Spearman’s rank, $\rho_{\rm{s}}=0.23$, $p=0.07$), but this is not surprising given that the masses of \tth{} clusters only span 0.7 orders of magnitude.

The accuracy of this classifier is substantially higher than existing methods, such as the phase-space definition of backsplash galaxies from \citetalias{rhee2017}. Applying that method to our simulated data gives a classification accuracy of $\perfEgrheAc\%$, and a backsplash sample whose purity and completeness are just $\calP=\perfEgrheBP\%$ and $\calC=\perfEgrheBC\%$. However, we note that the full implementation of our model uses 14 quantities to make this classification, as opposed to the simple phase-space information used by methods like \citetalias{rhee2017}; a more like-for-like comparison is described in \Sec{sec:simplest}, and the following subsections.

As expected, these are the purest and most complete samples produced by any implementation of our model. However, a substantial amount of data is required to use the model in this way, and measuring all 14 of these parameters for multiple galaxies would require extremely high-quality observations. In the following subsections, we analyse the results of our model in more detail, utilising only select subsets of the input parameters. This gives a more realistic expectation of how the model would perform, given a set of measured galaxy properties.

\subsection{Simplest model implementation}
\label{sec:simplest}

\begin{figure*}
\includegraphics[width=\textwidth]{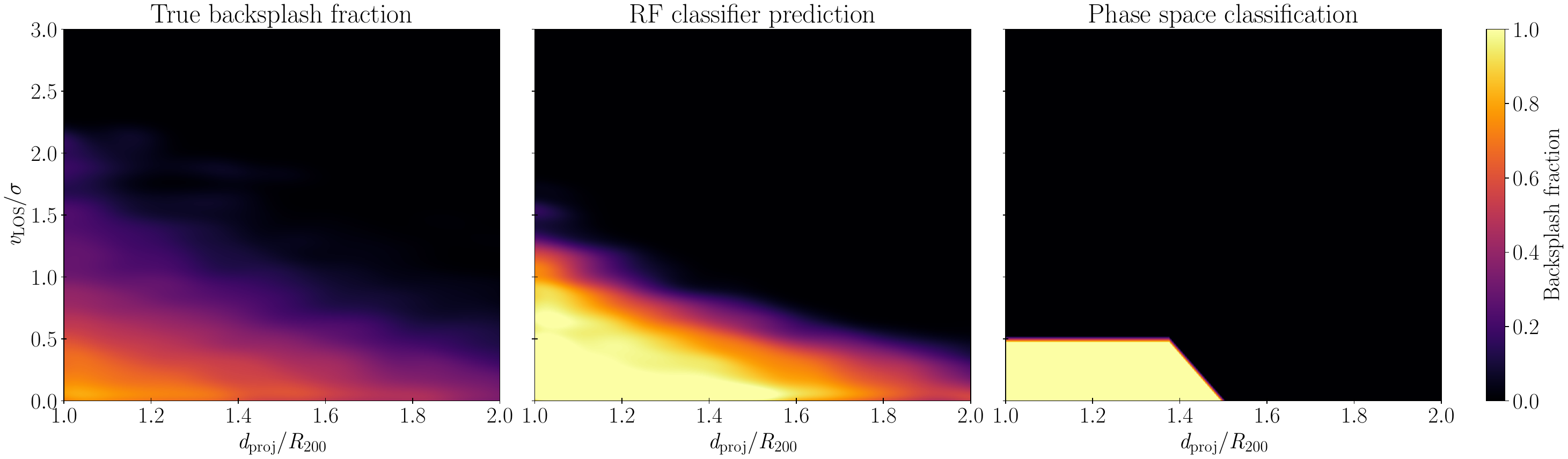}
\caption{Backsplash fraction as a function of projected distance from the cluster centre, $d_{\rm{proj}}$, in units of the cluster radius, $R_{200}$, and line-of-sight velocity relative to the cluster centre, $v_{\rm{LOS}}$, in units of the line-of-sight cluster velocity dispersion, $\sigma$. Left panel shows the true fraction of backsplash galaxies in our training set. Centre panel shows the fraction from our test set, predicted using the random forest classifier we develop in this work; this instance uses the minimal classifier, trained only on projected distance, line-of-sight velocity, and cluster magnitude gap. Right panel shows an example of an alternative classification method, based on regions of phase space. The classification shown comes from \citetalias{rhee2017}, but any of those displayed in \Fig{fig:phase_space_boundaries} would give a similar result. The left and centre panels  show a kernel density estimation (KDE) of the backsplash fraction, and the right panel shows the backsplash fraction imposed by the phase-space classification rule.}
\label{fig:phase_space_comp_basic}
\end{figure*}

The simplest version of our model, using only the projected cluster-centric distances, line-of-sight velocities, and $R$-band magnitude gap, allows galaxies to be classified with an accuracy of $\perfBasicAc\%$. This is very similar to the accuracy obtained by the phase-space definition of \citetalias{rhee2017}, which also gives an accuracy of $\perfEgrheAc\%$ when applied to our test data, for galaxies in the projected annulus between \radrange. However, the sample of backsplash galaxies selected by the random forest classifier has a completeness of $\perfBasicBC\%$, compared to just $\perfEgrheBC\%$ using \citetalias{rhee2017}, with only a marginally lower purity ($\perfBasicBP\%$, compared to $\perfEgrheBP\%$ from \citetalias{rhee2017}). We show the backsplash fraction estimated by our model as a function of projected position and line-of-sight velocity in \Fig{fig:phase_space_comp_basic}. The converse sample generated by the random forest classifier (the `infalling' galaxies) has a purity and completeness of $\calP=\perfBasicIP\%$ and $\calC=\perfBasicIC\%$ respectively, compared to $\calP=\perfEgrheIP\%$ and $\calC=\perfEgrheIC\%$ for the infalling galaxies selected by the \citetalias{rhee2017} classification. This performance by the random forest classifier uses a classification threshold p(Backsplash) of \perfBasicTh, calculated using the methodology described in \Sec{sec:threshold}.

In this simplest version of our model, we note that almost all galaxies in the bottom-left of this plot are assigned as backsplash, despite the fact that the true backsplash fraction peaks at about $70\%$. This comes from the fact that galaxies are treated on a case-by-case basis in our model. If every galaxy is assigned a $70\%$ chance of being a backsplash galaxy, then every one of them will be above the backsplash classification threshold. Consequently, every galaxy will be assigned to the backsplash class, while none of them will be assigned as infalling galaxies, resulting in $\sim100\%$ of the galaxies in this region being assigned as `backsplash'. We discuss this in further detail in \App{sec:overestimated}, and also show an alternative version of \Fig{fig:phase_space_comp_basic} that displays the average backsplash probability as a function of phase space.

The boundary between the backsplash and infalling galaxies selected through our method is blurred. This is in contrast to the hard boundary imposed by a phase-space classifier, which assumes that all galaxies within this region are backsplash, and all galaxies outside of this region are first-time infallers. This is primarily  due to the additional information included by the magnitude gap, which is a proxy for the dynamical state of each cluster. Dynamically disturbed clusters have fewer backsplash galaxies in their outskirts \citep{haggar2020}, and their backsplash galaxies are also concentrated around $R_{200}$ of the cluster, rather than extending to greater radii \citep{kuchner2022}. Consequently, the region of phase space containing backsplash galaxies is different for each cluster, giving these blurred boundaries when the data for many clusters are stacked. Some previous studies have employed similar methods -- for example, \citet{yoon2017} combine 2D positions and line-of-sight velocities with the \HI\ content of galaxies in the Virgo cluster to identify backsplash galaxies. They therefore do not find hard boundaries in the distribution of their backsplash galaxies in phase space. Our random forest method also allows a more complex, empirical division between backsplash and infalling galaxies to be determined, driven by the data, rather than by a theoretical prediction of the `backsplash region'. 

\begin{figure*}
\includegraphics[width=\textwidth]{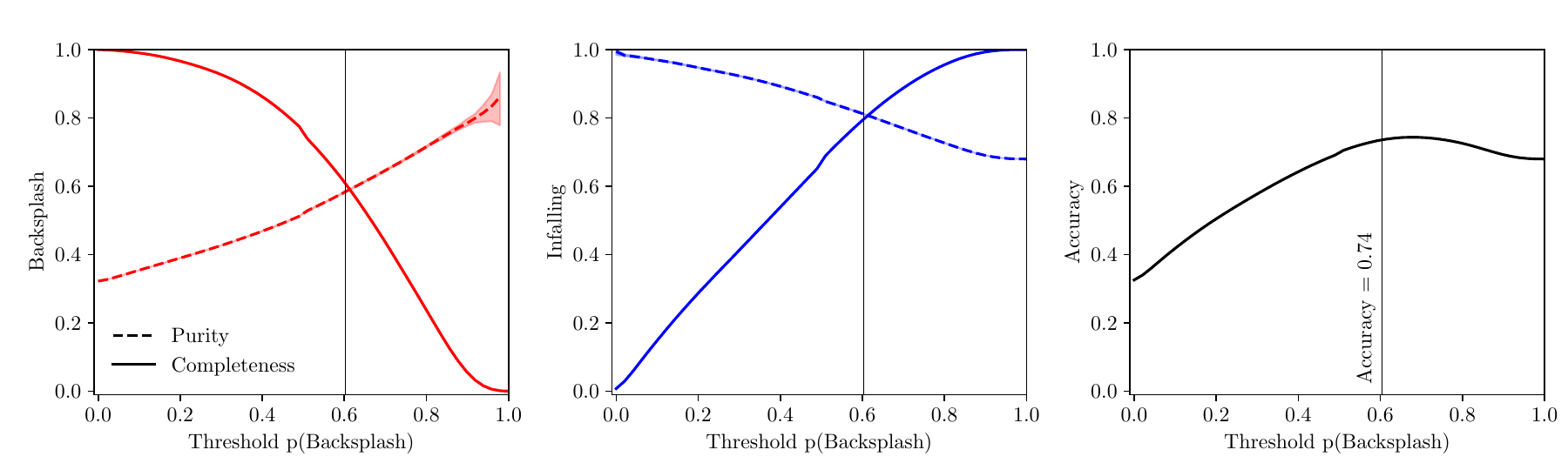}
\caption{Purity and completeness of backsplash (left) and infalling (centre) samples, as a function of backsplash galaxy threshold value. Right panel shows the overall accuracy of the model as a function of the threshold value. Vertical line represents the fiducial threshold value, $\rm{p(Backsplash)}=\perfBasicTh$, as described in \Sec{sec:threshold}. Shaded regions are the uncertainties derived by bootstrapping from a binomial distribution.}
\label{fig:prob_backsplash_basic}
\end{figure*}

As a result of this, we can manually adjust the threshold at which a galaxy is assigned as a backsplash galaxy, changing it from its default value of \perfBasicTh, chosen through the method described in \Sec{sec:threshold}. Doing this allows us to increase the purity of a sample at the cost of completeness, or vice versa. This is demonstrated in \Fig{fig:prob_backsplash_basic}, which shows how the purity ($\calP$) and completeness ($\calC$) of these samples, as well as the overall accuracy, vary for this basic implementation of the model. Increasing the threshold for assigning a backsplash galaxy to $0.82$ results in a high-completeness infalling sample ($\calP=72\%$; $\calC=97\%$), and a high-purity backsplash sample, albeit with a low completeness ($\calP=75\%$; $\calC=18\%$). Conversely, reducing this threshold to $0.42$ gives a high-purity infalling sample ($\calP=89\%$; $\calC=57\%$), and a high-completeness sample of backsplash galaxies ($\calP=48\%$; $\calC=85\%$). In both of these cases, the overall accuracy of the model remains above $65\%$. Different applications of this model would make use of either a highly-complete or highly-pure sample of galaxies, allowing the model to be applied to multiple different use cases. A user can manually adjust this threshold through the web app version of this classifier.

We also stress that these somewhat moderate values of $\calP$ and $\calC$ apply to the simplest version of our model, using only very basic galaxy properties. In the following subsections, we include further properties to improve the efficacy of our model.

\subsection{Addition of second velocity component direction}
\label{sec:velocity_direction}

Adding in the direction of each galaxy's velocity in the plane of the sky, $\theta_{\rm{v}}$, improves the accuracy of our method substantially. We note that this is not the magnitude of this velocity component, but simply the direction of a galaxy's present-day velocity projected in the plane of the sky, relative to the centre of the cluster (defined as the location of the central galaxy). This information is available indirectly, through observations of outflowing material from galaxies. Bent radio jets \citep{devos2021, koribalski2024} and ram pressure stripped tails \citep{jaffe2018, roberts2021} both allow the direction of a galaxy's velocity in the plane of the sky to be measured, assuming that it is travelling at a constant velocity through a static medium.

\begin{figure*}
\includegraphics[width=\textwidth]{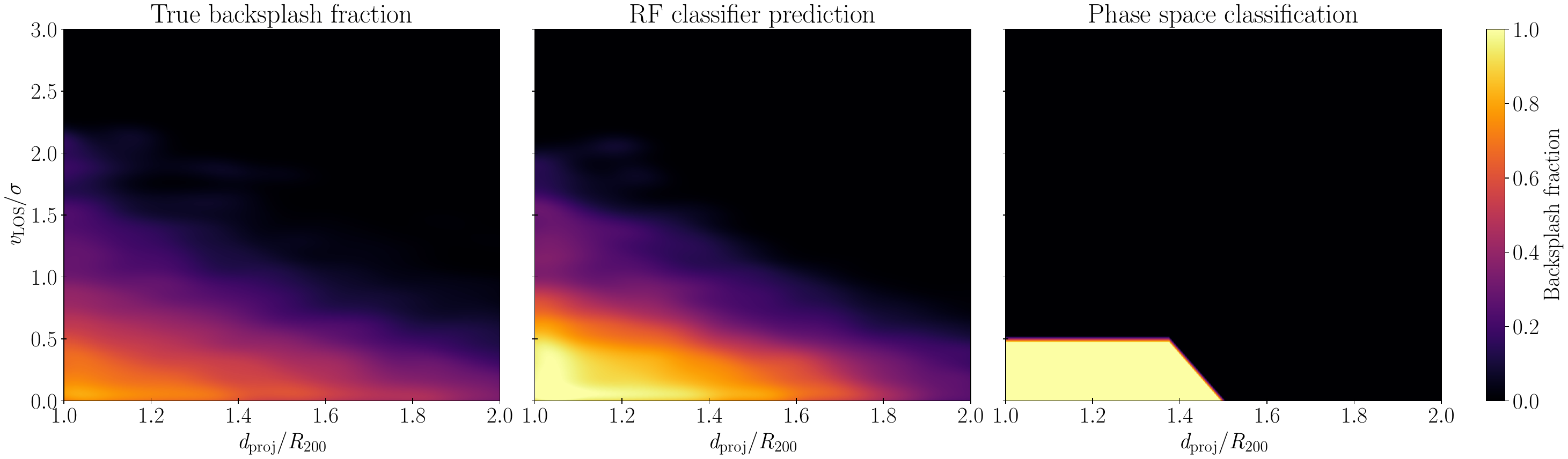}
\includegraphics[width=\textwidth]{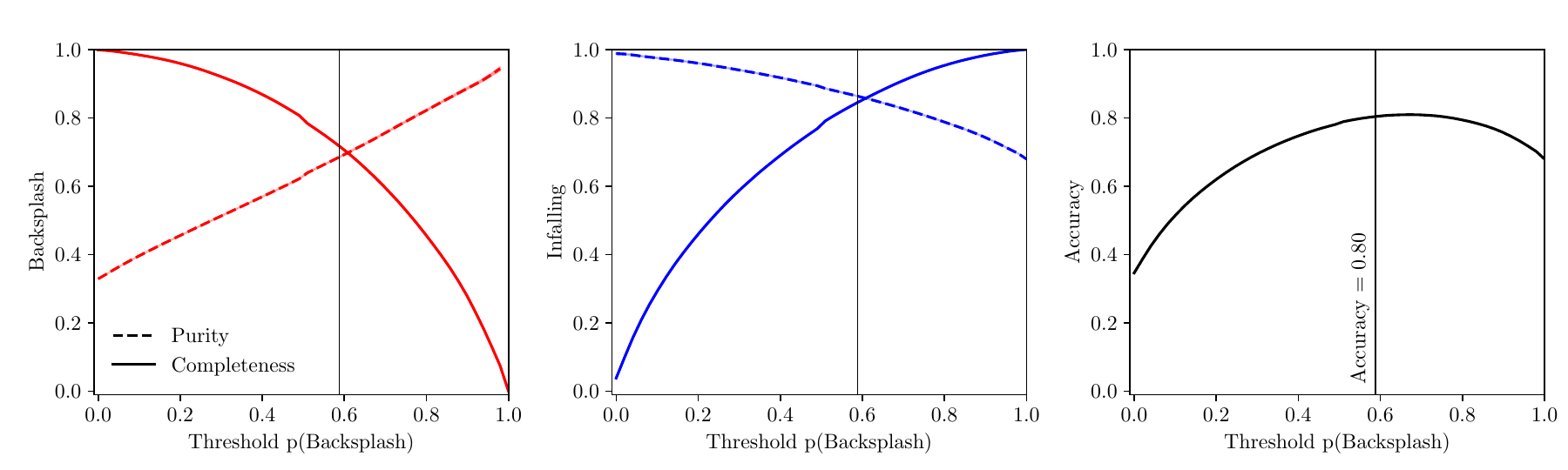}
\caption{Equivalent figures to \Fig{fig:phase_space_comp_basic} (top row) and \Fig{fig:prob_backsplash_basic} (bottom row), with the addition of the velocity direction projected in the plane of the sky, $\theta_{\rm{v}}$. Top row shows the backsplash fraction as a function of position-velocity phase space for the true sample of simulated backsplash galaxies (left), those identified by our classifier (centre), and those selected with a hard phase-space cut (right). Bottom row shows the purity and completeness of the produced backsplash (left) and infalling galaxy (centre) samples, and the accuracy of the model (right).}
\label{fig:extra_velocity}
\end{figure*}

Adding in this fourth variable, the accuracy of our model increases from $\perfBasicAc\%$ to $\perfVelocAc\%$. The produced backsplash galaxy sample has a purity and completeness of $\calP=\perfVelocBP\%$ and $\calC=\perfVelocBC\%$, while the sample of infalling galaxies has $\calP=\perfVelocIP\%$ and $\calC=\perfVelocIC\%$, a substantial improvement over existing methods. The plots equivalent to \Fig{fig:phase_space_comp_basic} and \Fig{fig:prob_backsplash_basic} are shown in \Fig{fig:extra_velocity}.

\subsection{Nearest neighbours and BCG information}
\label{sec:substructure}

Including three measures describing the local density around each galaxy ($d_{\rm{proj,nn4}}$) and the shape of the cluster (cos$(\phi_{*})$ and $e^{*}_{\rm{a,BCG}}$), makes a small improvement to the model, compared to the simplest version of the model (\Sec{sec:simplest}). The purity and completeness of the predicted backsplash galaxy sample changes from $\calP=\perfBasicBP\%$, $\calC=\perfBasicBC\%$, to $\calP=\perfNeighBP\%$, $\calC=\perfNeighBC\%$. The equivalent values for the infalling sample change from $\calP=\perfBasicIP\%$, $\calC=\perfBasicIC\%$, to $\calP=\perfNeighIP\%$, $\calC=\perfNeighIC\%$. 

The parameter importance measures for the six input parameters reveal that all of this improvement is due to the inclusion of the distance to the fourth nearest neighbour galaxy; the inclusion of information relating to the orientation of the central galaxy's major axis gives no significant improvement to the model. The additional classifying power provided by the distance to the fourth nearest neighbour is likely a consequence of the fact that backsplash galaxies are less likely to exist in small, bound groups, as these are tidally disrupted on passing through a cluster \citep{taylor2004, vandenbosch2018, haggar2023}. Indeed, in our simulations we find that, among galaxies in the projected radial range \radrange, $19\%$ of infalling galaxies have four neighbouring galaxies within a distance of $0.1R_{200}$\footnote{We emphasise again that this is the $R_{200}$ radius of the central cluster, not the galaxies in question. This distance is equal to approximately $250~\mathrm{kpc}$ for the cluster mass range we use in this work.}. This fraction is only $9\%$ for backsplash galaxies, demonstrating that backsplash galaxies are less likely to exist in these small `sub-clumps' around clusters. 

For brevity, we do not include an analogous figure to \Fig{fig:extra_velocity} for this iteration of the classifier, as the performance is very similar to that shown in \Fig{fig:phase_space_comp_basic} and \Fig{fig:prob_backsplash_basic}. However, in \App{sec:selection_criteria}, we give a complete demonstration of how this iteration of the model identifies backsplash galaxies, including the information in the top-central panel of \Fig{fig:extra_velocity}.

\subsection{Morphological properties of galaxies}
\label{sec:morphology}

Including the morphological parameters in our simulation (cos$(\phi_{*})$, $e^{*}_{\rm{a}}$, $a_{\rm{hl}}$, and $b_{\rm{hl}}$), plus the stellar mass of the galaxies, $m_{*}$, also has only a small impact on the model, marginally less so than including the distance to the fourth nearest neighbour galaxy. Specifically, the purity and completeness of the predicted backsplash galaxy sample are $\calP=\perfMorphBP\%$, $\calC=\perfMorphBC\%$, and the equivalent values for the infalling sample are $\calP=\perfMorphIP\%$, $\calC=\perfMorphIC\%$.

It is clear that the model benefits only slightly from the inclusion of this morphological information. Additionally, classifying backsplash and infalling galaxies based on their morphologies would mean that any subsequent analysis of the morphologies of these populations would be circular. Other properties of these samples (for example, their gas contents or AGN activity) could still be analysed and compared. However, given the additional complexity of measuring morphological properties of galaxies, we conclude that there are limited practical reasons for a user to provide this information in the model in its current state.

As described in \Sec{sec:complete}, it is likely that the low importance in these morphological parameters is due to the limited resolution of the simulations, which means that these galaxy-scale properties are noisy. In the future, we plan to re-train this model on a higher-resolution version of \tth{} simulations, once this data is available. Similarly to \Sec{sec:substructure}, we have chosen not include an analogous figure to \Fig{fig:extra_velocity} for this iteration of the classifier.
\\\\
We summarise the effectiveness of all these iterations of our model in \Tab{tab:accuracy}. Of each of the iterations of our model, the full classifier (\Sec{sec:complete}) performs best on all metrics, as expected.

\begin{table*}
	\centering
	\caption{Summary of results in \Sec{sec:results}. The accuracy of each of our five test cases is shown, and the purity, $\calP$, and completeness, $\calC$, of the samples of backsplash and infalling galaxies they produce. Also included is the optimal classification threshold, as defined in \Sec{sec:threshold}, to achieve this performance. For comparison, the performance of two existing methods based on phase-space information are included, one with a restrictive definition of backsplash galaxies \citep{rhee2017}, and one with a less restrictive definition \citep{ferreras2023}. We again note that the random forest model presented in this work can be controlled to be more or less restrictive, to produce samples of galaxies with a greater purity and lower completeness, or vice-versa.}
	\label{tab:accuracy}
    \begin{tabular}{l | c | c | c | c | c | c | c}
    \hline
    Model & Accuracy (\%) & Backs. $\calP$ (\%) & Backs. $\calC$ (\%) & Inf. $\calP$ (\%) & Inf. $\calC$ (\%)  & Class. threshold \\
    \hline
    Simplest (\ref{sec:simplest}) & \perfBasicAc & \perfBasicBP & \perfBasicBC & \perfBasicIP & \perfBasicIC & \perfBasicTh \\
    Simplest + neighbours (\ref{sec:substructure}) & \perfNeighAc & \perfNeighBP & \perfNeighBC & \perfNeighIP & \perfNeighIC & \perfNeighTh \\
    Simplest + morphology (\ref{sec:morphology}) & \perfMorphAc & \perfMorphBP & \perfMorphBC & \perfMorphIP & \perfMorphIC & \perfMorphTh \\
    Simplest + velocity angle (\ref{sec:velocity_direction}) & \perfVelocAc & \perfVelocBP & \perfVelocBC & \perfVelocIP & \perfVelocIC & \perfVelocTh \\
    Full RF classifier (\ref{sec:complete}) & \perfComplAc & \perfComplBP & \perfComplBC & \perfComplIP & \perfComplIC & \perfComplTh \vspace{2pt}\\
    \multicolumn{6}{l}{\textit{Existing methods:}\vspace{1pt}} \\
    \citet{rhee2017} & \perfEgrheAc & \perfEgrheBP & \perfEgrheBC & \perfEgrheIP & \perfEgrheIC & -- \\
    \citet{ferreras2023} & \perfEgferAc & \perfEgferBP & \perfEgferBC & \perfEgferIP & \perfEgferIC & -- \\
    \hline
    \end{tabular}
\end{table*}

\section{Proof-of-concept applications}
\label{sec:applications}

This model has been trained on simulated data, but is designed to be applied to observations. In this section, we discuss the potential applications of this work.

\subsection{Mock galaxy cluster from The300 simulations}
\label{sec:mock}

\begin{figure*}
\includegraphics[width=\textwidth]{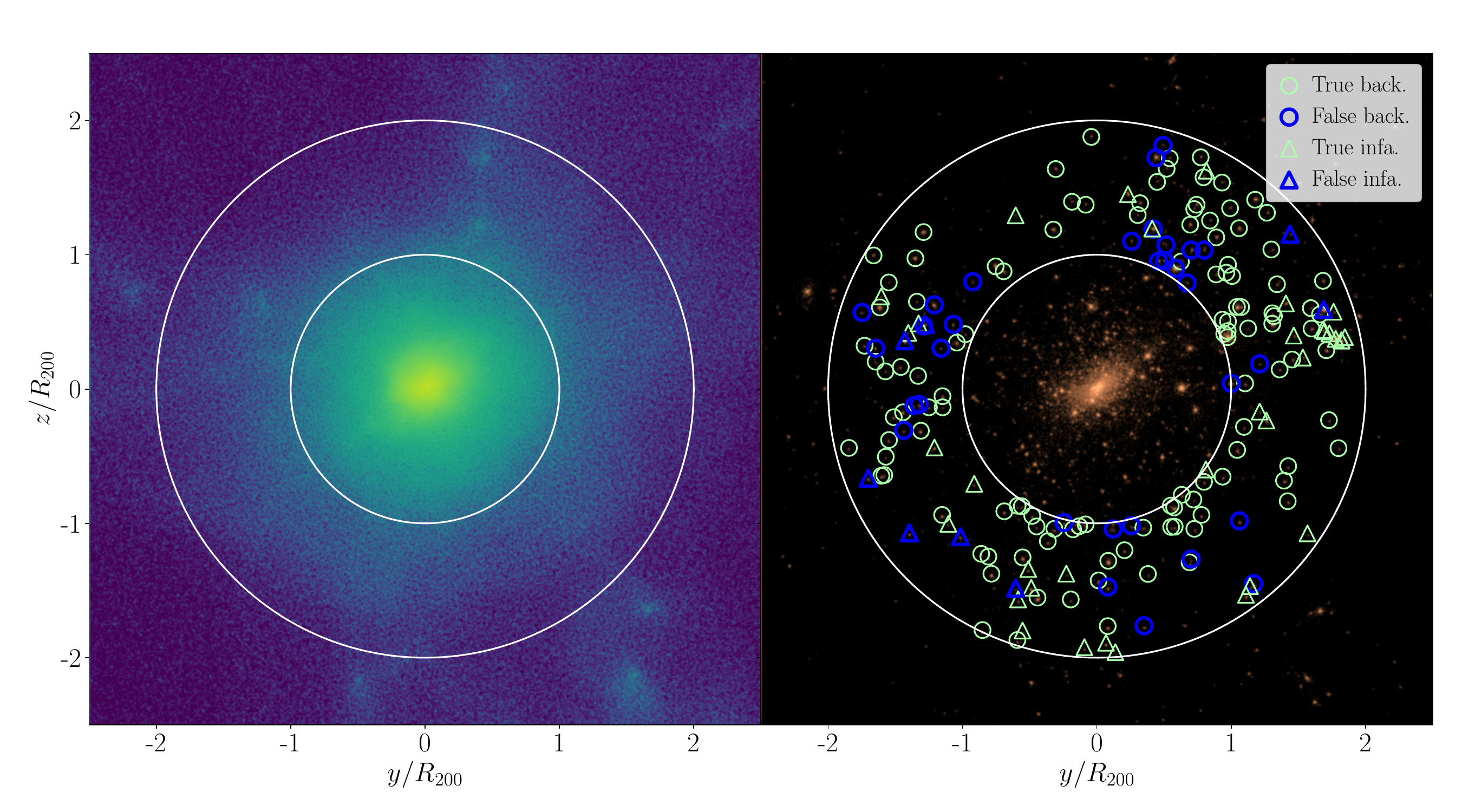}
\caption{Example cluster from \tth{} simulations (ID: {\sc{cluster\_0141}}). Left panel shows the mass distribution of gas in this cluster, right panel shows the distribution of stars. The large white circles represent distances of $R_{200}$ and $2R_{200}$ from the cluster centre. The right panel shows backsplash galaxies and infalling galaxies identified by our model, along with their status as true/false backsplash/infalling galaxies. Circles represent galaxies the model identifies as backsplash, and triangles are galaxies identified as first-time infallers. Light green markers represent correct predictions, dark blue markers are incorrect predictions. For example, a dark blue triangle is a backsplash galaxy that the model incorrectly identifies as an infalling galaxy -- a `false infaller'.}
\label{fig:example_cluster}
\end{figure*}

In \Fig{fig:example_cluster}, we show one simulated cluster from \tth{} simulations, as a demonstration of the galaxy samples it is possible to build using mass-complete spectroscopic redshifts of galaxies in the outskirts of clusters. These kinds of datasets will be available thanks to galaxy cluster surveys like the WHT Enhanced Area Velocity Explorer (WEAVE) Wide-Field Cluster Survey \citep[WWFCS; ][]{jin2024}, and the CHANCES survey with the 4-metre Multi-Object Spectroscopic Telescope \citep[4MOST; ][]{sifon2025}, as well as large spectroscopic surveys like \textit{Euclid} \citep{euclid2025}. To produce this, we re-ran our model, training it on all the simulated clusters except for one (\tth{} ID: {\sc{cluster\_0141}}, a dynamically relaxed cluster). We then applied it to this cluster, providing only $d_{\rm{proj}}$, $v_{\rm{LOS}}$, and $d_{\rm{proj,nn4}}$ for each galaxy, plus the magnitude gap of the cluster.

The resulting figure, \Fig{fig:example_cluster}, gives a visual representation of the expected performance of this model. Light green shapes show correct classifications by the model, and dark blue are incorrect classifications -- clearly this information would not be available for a real cluster, and so we have not used it in our analysis, besides for illustrative purposes.

Circled galaxies are those classified as backsplash galaxies, and triangles are galaxies identified as infallers. Applying this work to catalogues of galaxies around observed clusters would allow similar samples of backsplash and infalling galaxies to be constructed and compared. For example, in this `mock' cluster observation, the median stellar masses of the constructed backsplash and infalling populations are the same: $\textrm{log}_{10}(m_{*}/h^{-1}\,M_{\odot})=9.9^{+0.7}_{-0.2}$ and $\textrm{log}_{10}(m_{*}/h^{-1}\,M_{\odot})=9.9^{+0.6}_{-0.3}$, respectively (median and $1\,\sigma$ spread). However, the halo-to-stellar mass ratios of the backsplash galaxies are slightly lower ($20^{+12}_{-6}$, compared to $26^{+15}_{-10}$), demonstrating that backsplash galaxies' haloes have been stripped during their first passage through a cluster. Similarly, the median gas mass of backsplash galaxies is zero, as the vast majority of them have had all of their gas removed. This is compared to $\textrm{log}_{10}(m_{\textrm{gas}}/h^{-1}\,M_{\odot})=9.5^{+0.4}_{-0.3}$ for the infalling galaxies, although we acknowledge that this difference is likely partially due to the overactive ram pressure stripping in many hydrodynamical simulations. Nevertheless, none of this information was provided when classifying these galaxies, showing the potential for this model to provide insights on the impact of a passage through a cluster on galaxy properties.

The same patterns are also seen when comparing the stellar mass, gas mass, and halo-to-stellar mass ratio of the true populations of backsplash and infalling galaxies around this cluster, rather than those identified by the model. These two respective populations have stellar masses of $\textrm{log}_{10}(m_{*}/h^{-1}\,M_{\odot})=9.9^{+0.7}_{-0.2}$ and $\textrm{log}_{10}(m_{*}/h^{-1}\,M_{\odot})=9.9^{+0.5}_{-0.3}$, halo-to-stellar mass ratios of $19^{+7}_{-5}$ and $30^{+12}_{-10}$, and gas masses of $m_{\textrm{gas}}/h^{-1}\,M_{\odot}=0\pm0$ and $\textrm{log}_{10}(m_{\textrm{gas}}/h^{-1}\,M_{\odot})=9.5^{+0.5}_{-0.2}$ (median and $1\,\sigma$ spread). 

\begin{figure*}
\includegraphics[width=\textwidth]{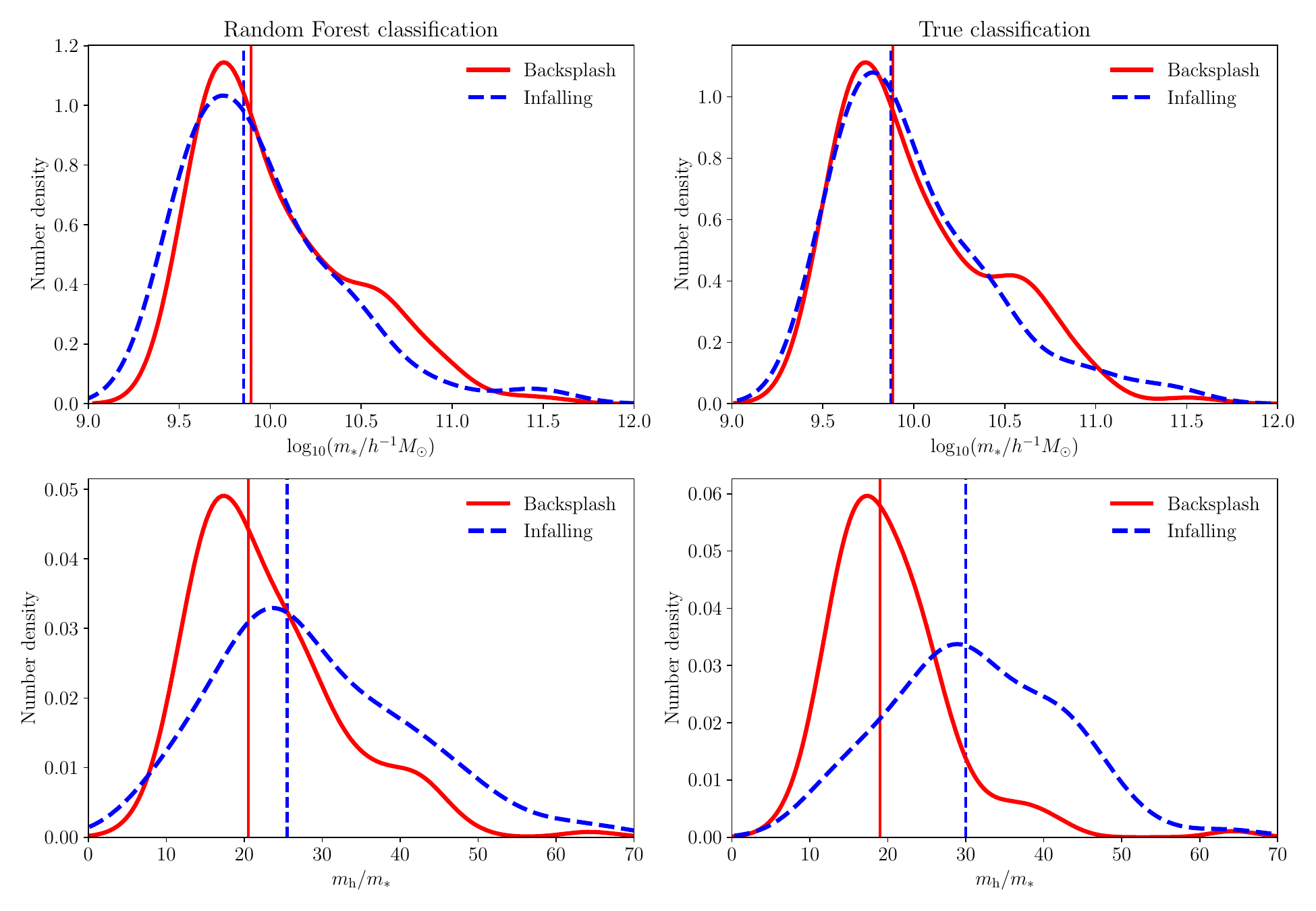}
\caption{Distributions of stellar masses (top), and halo-to-stellar mass ratios (bottom), for galaxies in the outskirts of our example cluster, shown in \Fig{fig:example_cluster}. Galaxies are separated into backsplash (solid red lines) and first-time infallers (dashed blue lines); this is based on their true classifications (right column), and the classification according to our model (left column). The median of each distribution is shown by the thin vertical line.}
\label{fig:example_cluster_mass_kdes}
\end{figure*}

\Fig{fig:example_cluster_mass_kdes} shows the distributions of the stellar masses, and the halo-to-stellar mass ratios, for the backsplash and infalling populations of galaxies around our test cluster. The left column shows the results for the populations identified by our classifier, and the right column shows the distributions for the true populations. As described above, the populations of infalling and backsplash galaxies around the test cluster have indistinguishable distributions of stellar masses, but different distributions in their halo-to-stellar mass ratios. The same is true of the populations of backsplash and infalling galaxies identified by our classifier, although some of the significance of the difference is lost. We note that the top row appears to show some galaxies with stellar masses below our lower stellar mass limit of $10^{9.5}~h^{-1}\,M_{\odot}$. This is due to the bandwidth used in the KDE calculation -- in practice, we exclude all galaxies below this limit from our analysis.

\subsection{Backsplash galaxies in the Virgo cluster}
\label{sec:virgo}

As a test case to demonstrate how this model can be applied to observational data, we next apply it to identify backsplash galaxies around the nearby Virgo Cluster. Specifically, we investigate whether ram pressure stripped candidates in the outskirts of Virgo (between \radrange) are preferentially falling into the cluster for the first time, or not. The positions, velocities, and magnitudes of galaxies in Virgo and its immediate outskirts have been well-studied and characterised in the public Virgo Cluster Catalogue \citep[VCC; ][]{binggeli1985} and Extended Virgo Cluster Catalogue \citep[EVCC; ][]{kim2014}. 

\HI\ emission from galaxies in the outskirts of Virgo has also been collected in the past, by the Virgo Cluster multi-Telescope Observations in Radio of Interacting galaxies and AGN (ViCTORIA) survey from the Low-Frequency Array \citep[LOFAR; ][]{edler2023}, and the VLA Imaging survey of Virgo galaxies in Atomic gas \citep[VIVA; ][]{chung2009}. We use this data to identify galaxies in the outskirts of Virgo (beyond $R_{200}$) that appear to have tails of trailing \HI\ gas. Assuming these to be ram pressure stripped galaxies, the direction of the tails can be taken to be antiparallel to the direction of each galaxy's velocity in the plane of the sky \citep[e.g.][]{bellhouse2017}. This allows us to calculate $(\theta_{\rm{v}})$, the angle between each galaxy's velocity and the cluster centre (the massive elliptical galaxy M87), shown in \Fig{fig:param_importance} to be highly valuable in identifying backsplash galaxies. While we note that Virgo also has a substantial second component \citep[`cluster B', centred around M49; ][]{boselli2023}, we do not consider this when determining the `centre' of Virgo, for consistency with our simulations.

Of the 62 galaxies between \radrange\ from the centre of Virgo with stellar masses greater than $10^{9.5}~M_{\odot}$, 14 have either LOFAR or VLA \HI\ maps. The directions of these \HI\ tails were identified by visual inspection; authors RH, ES, and CRM independently measured the angle between the \HI\ tail and the centre of Virgo for each galaxy, and the median value was taken as the `true' angle of the tail. Galaxies where two or more of the authors were unable to identify a clear direction of the tail were excluded from the sample; consequently, this sample contains only galaxies with strongly asymmetric  \HI\ distributions. Our final sample contains 10 galaxies between \radrange\ from the centre of Virgo, with well-defined orientations of their \HI\ tails; these galaxies are listed in \Tab{tab:virgo_galaxies}.

In total, we explicitly provide five inputs to the model ($d_{\rm{proj}}$, $v_{\rm{LOS}}$, $M_{12}$, $\theta_{\rm{v}}$, and $d_{\rm{proj,nn4}}$). We take an $R_{200}$ radius of Virgo of $0.97~\mathrm{Mpc}$ \citep{simionescu2017}, and a velocity dispersion of $799~\mathrm{km}\,\mathrm{s}^{-1}$ \citep{boselli2014}. These values actually place Virgo below the minimum halo mass of clusters used in our training set. While this potentially adds some bias to our classification, the impact should be minimal; Virgo is still a rich cluster of galaxies, and the five inputs used in this case have all been re-scaled to the cluster radius or velocity dispersion. We use an R-band magnitude gap of $0.3\,$mag \citep{kim2014}.

\begin{table*}
	\centering
	\caption{Names and properties of ten galaxies in the outskirts (between \radrange) of the Virgo cluster, with identified \HI\ tails. The Virgo Cluster Catalogue \citep[VCC; ][]{binggeli1985} and NGC IDs of these galaxies are listed. Final column also shows the label of each galaxy used in \Fig{fig:virgo_phase_space}.}
	\label{tab:virgo_galaxies}
    \begin{tabular}{ c | c | c | c | c | c | c | c }
    \hline
    VCC & NGC & $d_{\rm{proj}}/R_{200}$ & $v_{\rm{LOS}}/\sigma$ & cos$(\theta_{\rm{v}})$, & $d_{\rm{proj,nn4}}/R_{200}$ & \Fig{fig:virgo_phase_space} label \\
    \hline
    89 & 4189 & 1.27 & 1.45 & -0.85 & 0.03 & A \\
    92 & 4192 & 1.44 & -1.37 & -0.74 & 0.12 & B \\
    167 & 4216 & 1.10 & -1.03 & -0.14 & 0.03 & C \\
    187 & 4222 & 1.08 & -0.91 & 0.60 & 0.07 & D \\
    307 & 4254 & 1.05 & 1.82 & 0.41 & 0.05 & E \\
    596 & 4321 & 1.17 & 0.77 & -0.31 & 0.05 & F \\
    865 & 4396 & 1.03 & -1.35 & 0.93 & 0.10 & G \\
    1554 & 4532 & 1.78 & 1.32 & 0.47 & 0.09 & H \\
    1555 & 4535 & 1.27 & 1.26 & 0.86 & 0.04 & I \\
    2066 & 4694 & 1.33 & 0.26 & -0.84 & 0.08 & J \\
    \hline
    \end{tabular}
\end{table*}

\begin{figure}
\includegraphics[width=\columnwidth]{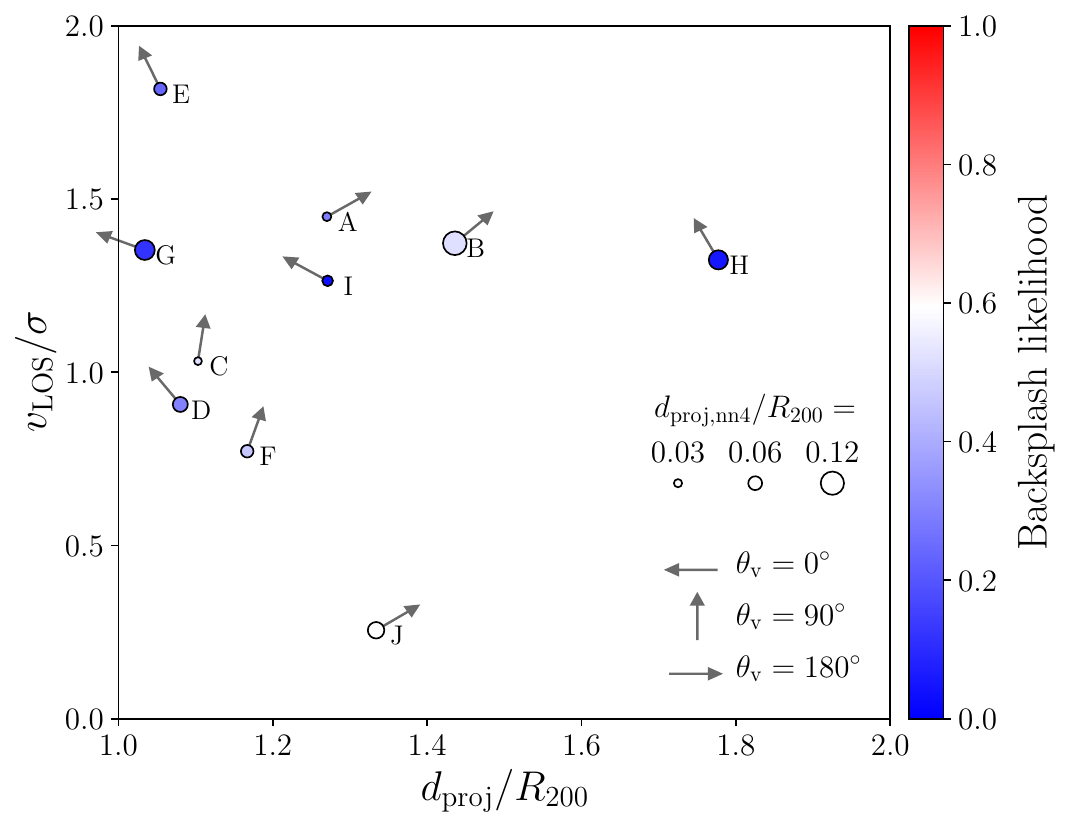}
\caption{Distribution of 10 galaxies with \HI\ tails in the outskirts of the Virgo cluster, in cluster-centric position-velocity space. The size of each point corresponds to $d_{\rm{proj,nn4}}$, such that larger points are more isolated from nearby galaxies. The arrows represent $\theta_{\rm{v}}$; arrows pointing left represent galaxies approaching the cluster centre, arrows pointing up show galaxies moving perpendicular to the cluster, and arrows pointing up show galaxies receding from the cluster centre. Each point is coloured by its likelihood of being a backsplash galaxy, as assigned by the random forest model. Every point is coloured blue, indicating the model predicts them to be infalling galaxies; darker shades of blue represent a greater likelihood that a galaxy is a first-time infaller. The labels (A-J) identify the 10 individual galaxies (see \Tab{tab:virgo_galaxies}).}
\label{fig:virgo_phase_space}
\end{figure}

\Fig{fig:virgo_phase_space} shows the backsplash likelihood metric that is calculated by the model for each of these galaxies, and the backsplash galaxy threshold value (black dashed line), which is equal to 0.59 in this instance. All 10 of the ram pressure stripped candidates in the outskirts of Virgo lie below this line, showing that the random forest classifier identifies all 10 of these as first-time infallers, not backsplash galaxies.

This is not a surprising result, as it is thought that ram pressure can remove substantial amounts of gas from a galaxy during a passage through a cluster \citep{mostoghiu2021, roberts2021}. Strongly asymmetric \HI\ tails imply that ram pressure stripping is in its early stages, when gas is being maximally removed from a galaxy. Our findings show that this stage does not continue throughout a galaxy's entire passage through a cluster, thereby supporting the idea that stripping of \HI\ gas preferentially happens during galaxies' first approach to a cluster. Several of these galaxies also appear to be located in dense substructures around Virgo. VCC 89, 92, 167, 187, and 307 are all located in the low-velocity cloud (LVC) region of Virgo, while VCC 1554 and 1555 are in the cluster B region, and VCC 2066 is very close to cluster C \citep{boselli2023}. All three of these structures are thought to be actively merging into Virgo, and simulations have shown that infalling groups like these are almost certainly approaching a cluster for the first time \citep{haggar2023}. Indeed, previous studies into the outskirts of Virgo \citep[e.g.][]{morgan2024, finn2025, morgan2025} have found galaxies in this region being actively quenched, that are consistent with a pre-processed population. Our results also suggest that these are a combination of pre-processed galaxies, and those being quenched by the outskirts of the cluster, rather than backsplash galaxies. 

Some of the classifications of individual galaxies are less conclusive than others; for example, Galaxy J (VCC 2066) lies very close to this classification threshold, indicated by its light colour in \Fig{fig:virgo_phase_space}, meaning the classification of this galaxy is inconclusive. This is primarily due to its low velocity relative to the cluster centre, and its \HI\ tail that is pointed towards the cluster centre (indicating a velocity vector away from the cluster). However, this galaxy also has three other galaxies within a projected distance of $75~\mathrm{kpc}$, indicating that it is likely a member of a dense, infalling structure, possibly cluster C. Indeed, if our analysis is repeated without the nearest neighbour information, VCC 2066 is instead classified as a backsplash galaxy. If the direction of the \HI\ tail is excluded from our analysis, all 10 of the galaxies are still classified as first-time infallers.

Galaxy B (VCC 92) is the most isolated of our sample, with its fourth nearest neighbour at a distance of $0.12R_{200}$ ($120~\mathrm{kpc}$). Despite that, it is still located in a densely-populated region, in the projected outskirts of the LVC group, which is thought to be falling into Virgo \citep{boselli2014}. The high line-of-sight velocity of VCC 92 also supports the argument that this is not a backsplash galaxy. Despite these two cases, it is still clear that our sample of 10 objects, detected in \HI\ and with asymmetric distributions of cold gas, are strongly biased towards galaxies approaching the cluster for the first time.

This result is an important proof of concept for the model we present in this work, demonstrating how this model, trained on simulations, can be applied to observational data from low-redshift clusters, to make predictions about the recent histories of galaxies in their outskirts.

\section{Conclusions}
\label{sec:conclusions}

In this work, we have developed a random forest classifier that can identify whether galaxies in the outskirts of massive clusters are approaching the cluster for the first time. This model has been trained on simulated data from \tth{} simulations, but using only quantities that can be measured observationally. This allows the model to be applied to observational data, to identify whether galaxies have previously experienced a cluster environment or not, on a case-by-case basis. We have also made this model available to the community in the form of a public web app\footnote{\url{\webappurl}}. This will be continually updated in the future as other galaxy properties are added to this model; we also plan to include more simulation data, including future iterations of \tth{} simulations with higher resolution and other physics models. More details about this web app are given in \App{sec:web_app}. Below, we summarise the key points of this paper:

\begin{itemize}
\item The model can be trained on-the-fly, using the galaxy properties available to the user to maximise classifying power. Galaxies are classified on a case-by-case basis, allowing separate samples of backsplash and infalling galaxies to be constructed. The dynamical properties of these galaxies are reasonably distributed, which is not the case for existing phase-space-based backsplash selection methods \citep[e.g.][]{muriel2014, rhee2017}, or previous approaches of quantifying the number of backsplash galaxies based on cluster properties \citep[e.g.][]{haggar2020}.

\item We have quantified the performance of the model with purity and completeness metrics. Using fairly simple quantities available from spectroscopic surveys of clusters (e.g. \Sec{sec:substructure}), a constructed sample of backsplash galaxies has a purity and completeness of $\calP=\perfNeighBP\%$, $\calC=\perfNeighBC\%$. The complementary sample of first-time infallers has $\calP=\perfNeighIP\%$, $\calC=\perfNeighIC\%$. 

\item These values can be improved further with the inclusion of additional information. In particular, increased classifying power is provided by the velocity direction in the plane of the sky ($\theta_{\rm{v}}$; \Sec{sec:velocity_direction}), available from observations like bent radio jets \citep{devos2021, koribalski2024} and ram pressure stripped tails \citep{jaffe2018, roberts2021}. The model has been built such that any combination of the observed properties in \Sec{sec:properties} can be provided, and the model will make the most accurate predictions based on those.

\item Additionally, the classification threshold can be manually tuned, allowing a user to be either more or less strict in the identification of backsplash galaxies. \Fig{fig:prob_backsplash_basic} and \Fig{fig:extra_velocity} show the associated performance of the model; these demonstrate how varying this threshold can be used to construct highly-complete or highly-pure samples of either infalling or backsplash galaxies.

\item In \Sec{sec:virgo}, a proof-of-concept study using this model has been carried out, by applying it to galaxies in the outskirts (\radrange) of the Virgo galaxy cluster. We selected galaxies with visible asymmetric distributions of \HI\ gas, indicating that they were actively being ram pressure stripped. All 10 of the galaxies we examined in Virgo are labelled as first-time infallers, despite some having some properties that might typically be associated with backsplash galaxies. This supports the idea that cold gas is strongly removed from a galaxy during its first approach to a cluster, and that little remains after a single passage through a dense cluster centre. 

\item We have published this model as a publicly-available web app (\url{\webappurl}). A user can provide a list of galaxy properties from \Sec{sec:properties} in the form of a {\sc{.csv}} file, and the platform will provide a backsplash likelihood for each galaxy, as well as some basic plots of the data. Detailed instructions for using the web app can be found in \App{sec:web_app}.
\end{itemize}

There are multiple potential uses for a model that can separate former cluster members from infalling galaxies. In the near future, we plan to apply this model to wider samples of galaxies, from more low-redshift clusters beyond just the Virgo Cluster. In particular, we plan to study how the star formation rates, and morphology, of infalling galaxies, cluster members, and backsplash galaxies differ. The ability of our model to construct highly-pure samples will be particularly useful for this, as it will allow us to compare representative populations of these different classes of galaxies. This will provide further insight to the timescales over which different environmentally-dependent mechanisms drive galaxy evolution.

In a follow-up study, we also plan to build on this model by extending this work to higher redshifts, and also by including other cosmic environments in our analysis. We have already begun to probe this by including a measure of local density and substructure ($d_{\rm{proj,nn4}}$). We plan to use more metrics like this to identify galaxies that have been pre-processed by smaller galaxy groups, not just by massive clusters in the form of backsplash galaxies. An upcoming study (Jordan et al., in prep.) will also use \tth{}  simulations to examine galaxies' group membership through machine learning methods. Other metrics of the dynamical state and structure of clusters can probe different intrinsic cluster properties and timescales \citep{haggar2024_ds,vallesperez2025}, which may also provide further information on galaxies' journeys through different cosmic environments.

\section*{Acknowledgements}

This work has been made possible by \threehun\ collaboration\footnote{\url{https://www.the300-project.org}}. This work has received financial support from the European Union's Horizon 2020 Research and Innovation programme under the Marie Sk\l{}odowskaw-Curie grant agreement number 734374, i.e. the LACEGAL project\footnote{\url{https://cordis.europa.eu/project/rcn/207630\_en.html}}. \threehun\ simulations used in this paper have been performed in the MareNostrum Supercomputer at the Barcelona Supercomputing Center, thanks to CPU time granted by the Red Espa\~nola de Supercomputaci\'on.

RH thanks Tomas Hough, and the members of the unofficial Waterloo `galaxies office', for productive discussions relating to this work. ES thanks the University of Waterloo for their support of the Canada Rubin Fellowship programme. CRM acknowledges support from an Ontario Graduate Scholarship. AK is supported by project PID2024-156100NB-C21 financed by MICIU /AEI/10.13039/501100011033 / FEDER, UE. He further thanks Red Hot Chili Peppers for under the bridge. WC gratefully thanks Comunidad de Madrid for the Atracci\'{o}n de Talento fellowship no. 2020-T1/TIC19882 and Agencia Estatal de Investigaci\'{o}n (AEI) for the Consolidaci\'{o}n Investigadora Grant CNS2024-154838. He further acknowledges the Project PID2024-156100NB-C21 financed by MICIU/AEI /10.13039/501100011033/FEDER, EU and ERC: HORIZON-TMA-MSCA-SE for supporting the LACEGAL-III (Latin American Chinese European Galaxy Formation Network) project with grant number 101086388 and the science research grants from the China Manned Space Project. JET acknowledges support from the Natural Sciences and Engineering Research Council of Canada (NSERC) through a Discovery Grant.

This work makes use of the {\sc{NumPy}} \citep{vanderwalt2011}, {\sc{SciPy}} \citep{virtanen2020}, {\sc{Matplotlib}} \citep{hunter2007}, {\sc{pandas}} \citep{mckinney2010}, {\sc{scikit-learn}} \citep{pedregosa2011}, and {\sc{Astropy}} \citep{astropy2022} packages for {\sc{Python}}\footnote{\url{https://www.python.org}}. This work also made use of GitHub Copilot\footnote{\url{https://github.com/features/copilot}} in the development of the accompanying web app. 

Preliminary results from this study, plus some of the text in \Sec{sec:rf}, were originally presented in the PhD thesis of RH, which can be accessed at: \url{https://eprints.nottingham.ac.uk/69844/}.

\section*{Data Availability}

The simulations underlying this work have been provided by \threehun\ collaboration. The data may be shared on reasonable request to the corresponding author, with the permission of the collaboration. The classifier is publicly-available as a web app (\url{\webappurl}).



\bibliographystyle{mnras}
\bibliography{main} 



\appendix

\section{User guide: Backsplash galaxy classifier}
\label{sec:web_app}

\begin{figure}
    \centering
    \includegraphics[width=\linewidth]{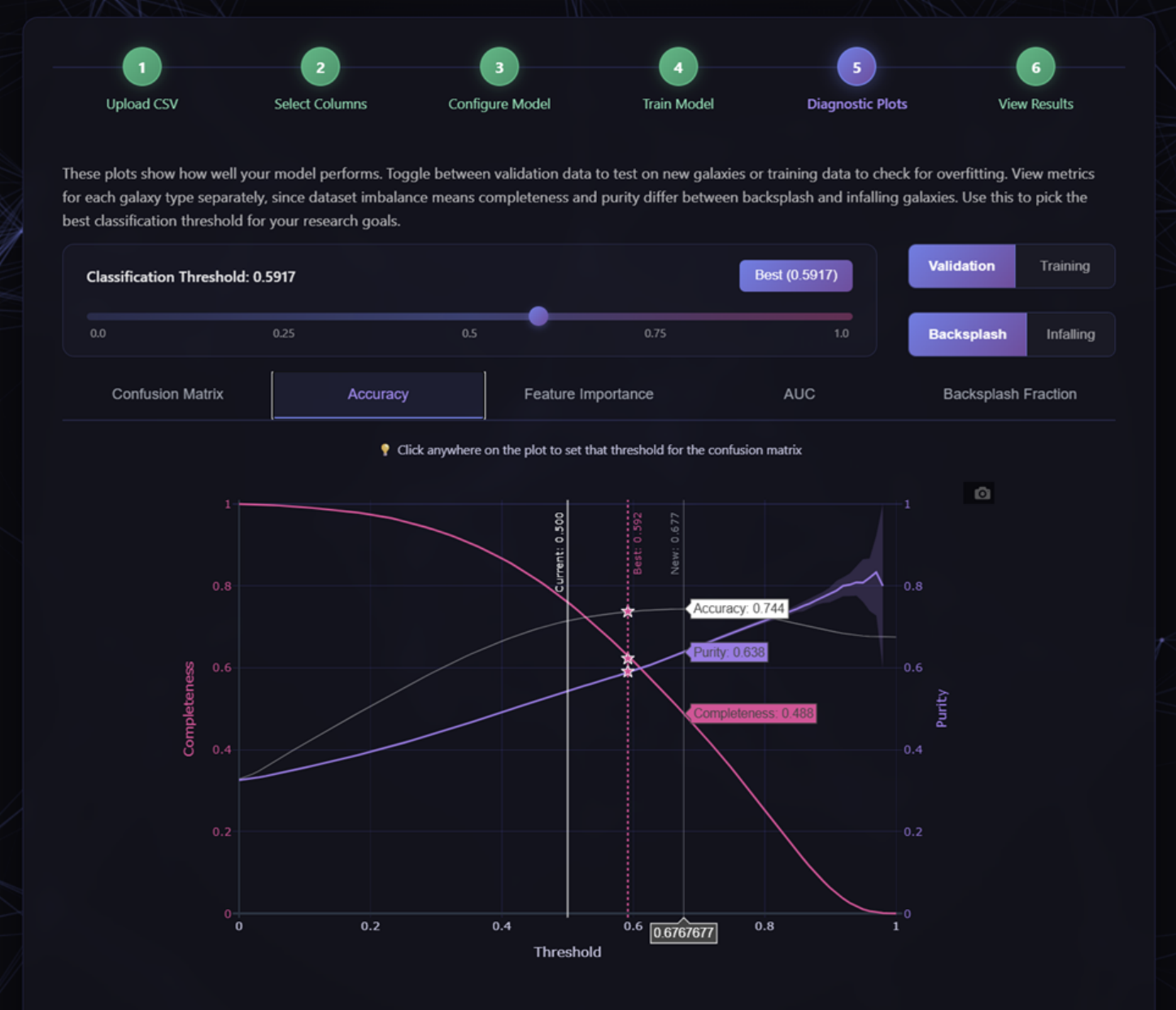}
    \caption{An example screenshot of the web app interface, showing the interactive step where the user can adjust the classification threshold to reach the desired purity or completeness.}
    \label{fig:webapp}
\end{figure}

In this Appendix, we provide a detailed guide on how to use the classifier presented in this work using the online web app\footnote{\url{\webappurl}}. The web app allows the users an easy access to our model, without requiring any local code installation or data download. A user can provide their own catalogue of galaxy properties, including a subset of observables they have available, and obtain backsplash classifications on-the-fly.

The classifier is trained on a cloud server using the training data from \tth{} simulations, as described in \Sec{sec:simulations}. This training catalogue is hosted on the server and is not accessible by the end-point user. The description of the columns in the training catalogue used by the classifier, given in \Sec{sec:properties}, along with their column names, is listed on the first page of the web app.

First, a user uploads the catalogue of their observed galaxies as a comma-separated file (CSV). After the session ends, all user-uploaded data is deleted. The columns in their file must contain some, but not necessarily all, columns used by the classifier (\Sec{sec:rf}). The column names do not have to match the training data: the web app then uses a fuzzy matching algorithm to assign the user's columns to those in the training catalogue. For example, columns labelled \texttt{log\_mstar}, \texttt{stellar\_mass}, or \texttt{lmass} would all map to the galaxy stellar mass column, \texttt{mstar}. The user is then asked to confirm that the column matching was accurate, and they can choose to reassign some columns to fix potential mismatches. While the column name mapping is flexible, it is crucial that the user-provided columns match the unit convention expected by the algorithm: stellar masses must be in units of $\log M/M_{\odot}$, the nearest neighbor distances in units of $R_{200}$, and so on, as described in \Sec{sec:properties}.

Next, the user can adjust the hyperparameters of the classifier: the number of trees in the forest, the size of the training subset, and the maximum depth. We recommend using the default options, as these are the parameters used in this paper. At this stage, the user can also enable permutation-based feature importances for a more robust importance estimation (see \Sec{sec:complete}), and choose how the `best' classification threshold is decided (\Sec{sec:threshold}). We recommend to keep the `best' threshold to most accurately predict the backsplash fraction in each training set cluster. The user has an opportunity to choose a different threshold after the model is trained.

The next step is to train the model. The model is trained on the subset of the 14 possible columns provided by the user. Additionally, there are several more possible quantities that can be provided by a user -- these are not described throughout this paper and are excluded from our analysis, as they are generally not feasible to measure in observations. Further galaxy properties will also be added over time as more simulation data becomes available, as described in \Sec{sec:conclusions}. Considering the subset of these galaxy properties in the user's dataset may vary, there are ${>10^{4}}$ possible combinations of the parameters to train on -- therefore, it is unfeasible to pre-train the model for all those options in advance. Instead, once the user has matched their columns to the training set, a new random forest classifier is trained using \tth{} data on-the-fly on the remote server. Since random forests are lightweight, this step takes under a minute. 

After the training step is complete, we provide the user with four diagnostic plots constructed using the training and validation data from \tth{}: the confusion matrix of the classifier (equivalent to \Fig{fig:confusion_all}), the feature importances (\Fig{fig:param_importance}), the purity and completeness diagnostic plots (\Fig{fig:prob_backsplash_basic}, \Fig{fig:extra_velocity}, and \Fig{fig:webapp}), and the `area under the curve' diagnostic as an alternative to evaluate the model performance. These plots do not include the user's data and are simply used to evaluate the performance of the trained model. Finally, we also provide the fraction of galaxies classified as backsplash in \tth{} clusters and the user's catalogue for comparison, and display a plot showing the backsplash fraction as a function of cluster magnitude gap in \tth.

At this stage, the user can interactively adjust the `classification threshold' slider to increase or decrease the p(Backsplash) threshold, described in \Sec{sec:threshold}, which dictates when a galaxy is assigned to the backsplash class. The diagnostic plots will update to reflect the new purity, completeness, and accuracy values. Once the user is satisfied with their choice, they can click `Continue to results' to run the final classification on their dataset. They can then download an augmented version of their catalogue with two new columns: the probability of a galaxy being a backsplash galaxy, and the determined class for each row. After the user ends their session, their data is deleted.

\section{Overestimated backsplash fraction in simple random forest model}
\label{sec:overestimated}

In \Sec{sec:simplest}, we noted that in the bottom-left of the phase-space diagram (\Fig{fig:phase_space_comp_basic}), almost $100\%$ of the galaxies with $d_{\textrm{proj}}\sim R_{200}$ and $v_{\textrm{LOS}}\sim 0$ are identified as backsplash galaxies, despite the `truth' panel showing that this is not the case. As described above, this is due to the fact that galaxies are treated on a case-by-case basis; if every galaxy is more likely to be a backsplash galaxy than an infalling galaxy, the estimated backsplash fraction in this region will be $100\%$, even though some of these galaxies are not in fact backsplash.

\begin{figure*}
\includegraphics[width=0.8\textwidth]{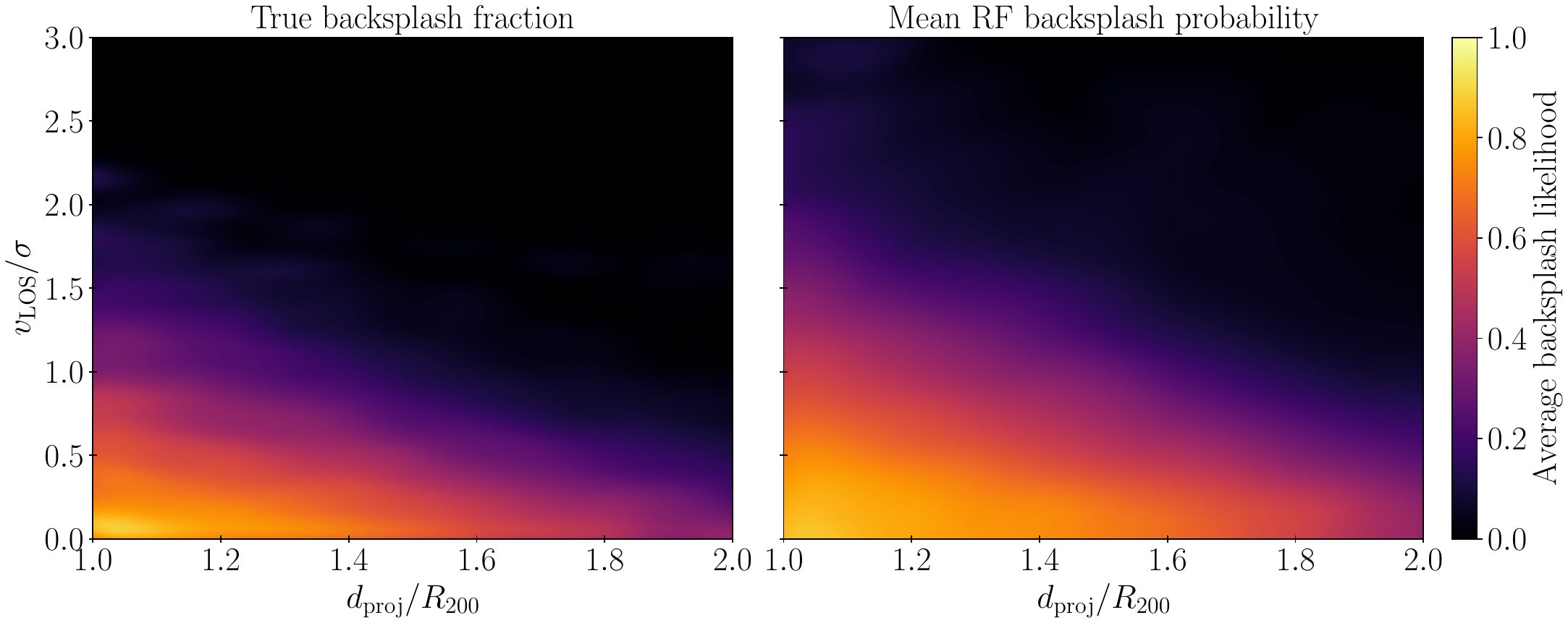}
\caption{Left panel: true fraction of backsplash galaxies in our simulated clusters, as a function of position-velocity phase space (as also shown in the left panel of \Fig{fig:phase_space_comp_basic}. Right panel: average likelihood of being assigned a backsplash galaxy by the random forest classifier (in the basic implementation, described in \Sec{sec:simplest}). The two distributions are similar, but this will still result in an overestimation of the backsplash fraction at low $d_{\rm{proj}}$ and low $v_{\rm{LOS}}$. We note that the same is also true for phase-space-based classifiers, which assign all galaxies in a specific region of phase space as backsplash.}
\label{fig:phase_space_alternative}
\end{figure*}

We illustrate this point with \Fig{fig:phase_space_alternative}. The left panel of this figure is identical to the left panel in \Fig{fig:phase_space_comp_basic}, but the right panel instead shows the average likelihood of being a backsplash galaxy, as a function of phase space. These two panels are far more closely aligned with one another; the average likelihood of being a backsplash galaxy assigned by the classifier matches well to the fraction of galaxies that are backsplash.

\section{Backsplash selection criteria using nearest neighbours and BCG properties}
\label{sec:selection_criteria}

\begin{figure*}
\includegraphics[width=0.85\textwidth]{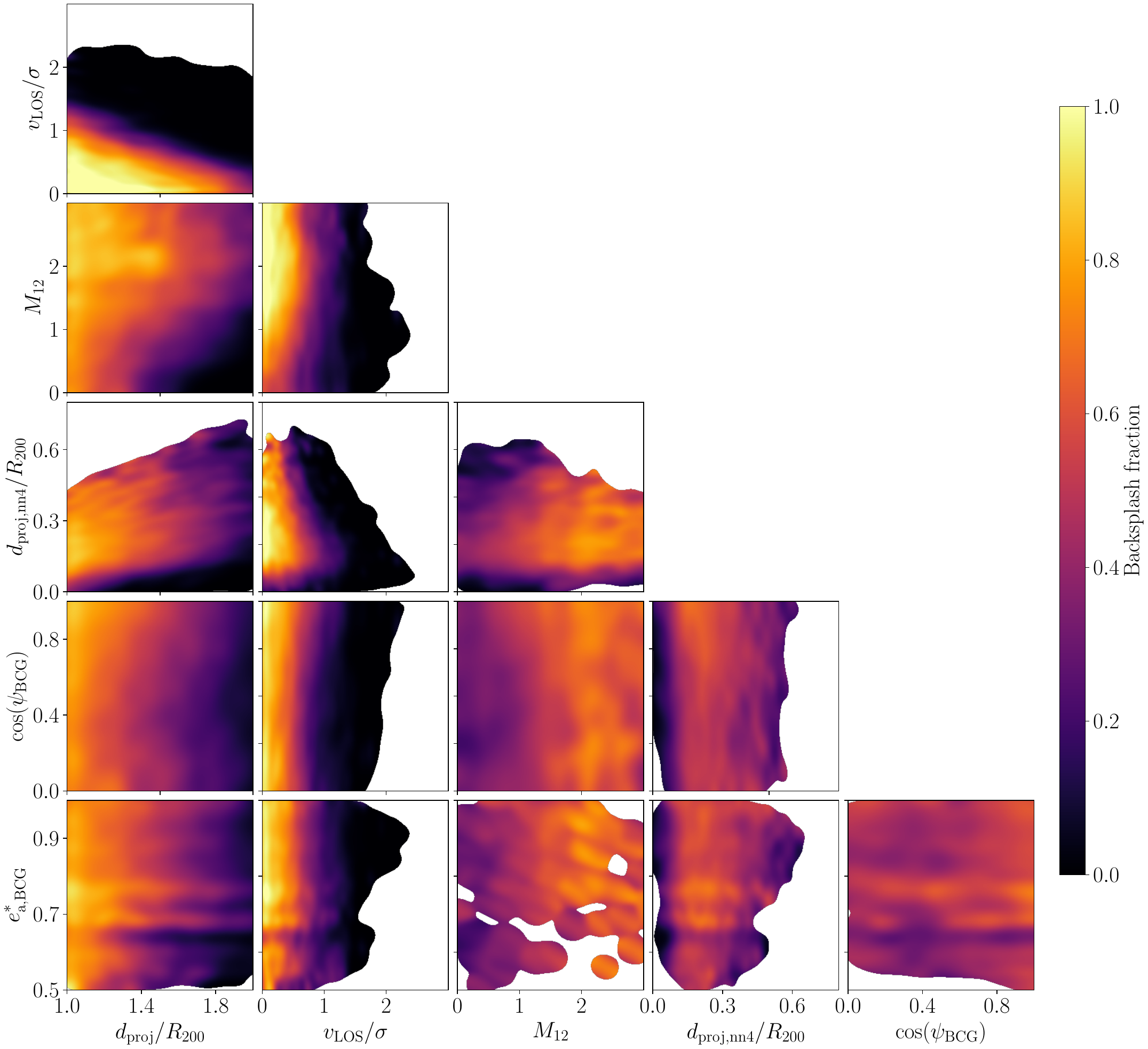}
\caption{Backsplash fraction as a function of six quantities used to train our random forest classifier in \Sec{sec:substructure}. This shows how likely galaxies are to be classified as backsplash by the model, giving an indication to how the decision trees are working. Lighter colours represent combinations of parameters that are more likely to lead to a galaxy being classified as `backsplash'. White regions show regions of each space with few galaxies (a total number density less than $1\%$ of the peak number density of galaxies).}
\label{fig:corner_plot}
\end{figure*}

In \Fig{fig:corner_plot}, we show how the version of our model described in \Sec{sec:substructure} selects backsplash galaxies. In each panel, the fraction of galaxies in the test clusters that are identified as backsplash is shown as a function of the six quantities used to train this iteration of the model: $d_{\rm{proj}}$, $v_{\rm{LOS}}$, $M_{12}$, $d_{\rm{proj,nn4}}$, cos$(\phi_{*})$, and $e^{*}_{\rm{a,BCG}}$. This demonstrates how backsplash galaxies are assigned by the random forest classifier, based on these six properties. We note that the top-left panel of this, showing the predicted backsplash fraction as a function of $d_{\rm{proj}}$ and $v_{\rm{LOS}}$, is exactly analogous to the central plot in \Fig{fig:phase_space_comp_basic}, or the top-centre plot in \Fig{fig:extra_velocity}.

Many of the patterns in this figure are broadly as expected. Again, backsplash galaxies are more likely to have low $d_{\rm{proj}}$ and $v_{\rm{LOS}}$ -- that is, be nearby to the cluster boundary, and have low velocities. There are also fewer backsplash galaxies around clusters with low magnitude gaps, $M_{12}$, which we interpret as dynamically disturbed clusters. There is a particularly strong deficit of backsplash galaxies at $d_{\rm{proj}}\gtrsim1.5R_{200}$ in disturbed clusters, consistent with previous findings \citep[e.g. Fig. 11 of][]{haggar2020}. Most galaxies with several nearby galaxies ($d_{\rm{proj,nn4}}<0.1R_{200}$) are first-time infallers, confirming that infalling galaxy groups are systematically less likely to contain backsplash galaxies. Again, this effect is stronger at distances of $d_{\rm{proj}}\gtrsim1.5R_{200}$ from the cluster. The classification of galaxies is not strongly dependent on the parameters $\rm{cos}(\psi_{\rm{BCG}})$ and $e^{*}_{\rm{a,BCG}}$, as also shown in \Fig{fig:param_importance}.

\bsp	
\label{lastpage}
\end{document}